\documentclass[12pt, draftclsnofoot, onecolumn]{IEEEtran}

\usepackage{booktabs} 

\usepackage{amsmath}
\usepackage{amssymb}
\usepackage{amsthm}
\usepackage{xfrac}
\usepackage{color}
\usepackage{graphicx}
\usepackage{stfloats}
\usepackage{algorithm}
\usepackage{algorithmic}
\usepackage{flushend}
\usepackage{caption}
\usepackage{enumitem}

\usepackage{amsmath}
\usepackage{amssymb}
\usepackage[normalem]{ulem}
\usepackage{graphicx}
\usepackage{flushend}
\usepackage{url}
\usepackage{color}
\usepackage{algorithm}
\usepackage{algorithmic}
\usepackage{ragged2e}
\usepackage{amssymb}
\usepackage{fancyvrb}
\usepackage[numbers,sort&compress]{natbib}

\newcommand{\mf}{\mathbf}
\newcommand{\mc}{\mathcal}
\newcommand{\mb}{\mathbb}

\newcommand{\bs}{\boldsymbol}

\usepackage{adjustbox}
\newlist{enumeratebyten}{enumerate}{1}
\setlist[enumeratebyten]{label={\textbf{Q\arabic*:}},ref={Q\arabic*:}}

\DeclareMathOperator{\argmax}{argmax}
\usepackage{colortbl}

\usepackage{booktabs,makecell,multirow,hhline} 
\usepackage{xcolor,colortbl,pgf}
\definecolor{Color1}{RGB}{207,224,252}
\definecolor{Color2}{RGB}{207,252,249}
\definecolor{Color3}{RGB}{189,175,247}

\ifCLASSINFOpdf
\else
\fi

\begin{document}

\newcommand{\todo}[1]{\textbf{\color{red}[XX #1 XX]}}

\title{\textit{DeepFIR:} Addressing the Wireless Channel Action in Physical-Layer Deep Learning}

\author{Francesco Restuccia, Salvatore D'Oro, Amani Al-Shawabka, Bruno Costa Rendon, Stratis Ioannidis, and Tommaso Melodia \vspace{-1.4cm} 
\thanks{The authors are with the Institute for the Wireless Internet of Things, Department of Electrical and Computer Engineering, Northeastern University, Boston, MA, 02215 USA. Corresponding author e-mail: melodia@northeastern.edu.}} 

\maketitle 

\vspace{-0.5cm}
\begin{abstract}
Deep learning can be used to classify waveform characteristics (\textit{e.g.}, modulation) with accuracy levels that are hardly attainable with traditional techniques. Recent research has demonstrated that one of the most crucial challenges in wireless deep learning is to counteract the channel action, which may significantly alter the waveform features. The problem is further exacerbated by the fact that deep learning algorithms are hardly re-trainable in real time due to their sheer size. This paper proposes \textit{DeepFIR}, a framework to counteract the channel action in wireless deep learning algorithms \textit{without retraining the underlying deep learning model}. The key intuition is that through the application of a carefully-optimized digital finite input response filter (FIR) at the transmitter's side, we can apply tiny modifications to the waveform to strengthen its features according to the current channel conditions. We mathematically formulate the \textit{Waveform Optimization Problem} (WOP) as the problem of finding the optimum FIR to be used on a waveform to improve the classifier's accuracy. We also propose a data-driven methodology to train the FIRs directly with dataset inputs. We extensively evaluate \textit{DeepFIR} on a experimental testbed of 20 software-defined radios, as well as on two datasets made up by 500 ADS-B devices and by 500 WiFi devices and a 24-class modulation dataset.  Experimental results show that our approach (i) increases the accuracy of the radio fingerprinting models by about 35\%, 50\% and 58\%; (ii) decreases an adversary's accuracy by about 54\% when trying to imitate other device's fingerprints by using their filters; (iii) achieves 27\% improvement over the state of the art on a 100-device dataset; (iv) increases by 2x the accuracy of the modulation dataset.\vspace{-0.3cm}
\end{abstract}

\section{Introduction}
The sheer number of wireless devices that will soon join the Internet of Things (IoT) will inevitably make the radio spectrum an extremely dynamic environment, where closed-form models will be difficult or impossible to formulate \cite{jagannath2019machine}. To address the severely increased spectrum complexity, the community is  using data-driven \textit{deep learning} techniques directly spun off of the computer vision domain  \cite{lecun2015deep}. 

Deep learning allows to capture relevant features in classification problems lacking tractable closed-form mathematical formulations. These models -- based on highly non-linear neural networks -- can address wireless problems such as signal/traffic classification \cite{OShea-ieeejstsp2018,Restuccia-infocom2019,o2016end},  radio fingerprinting \cite{Sankhe-infocom2019,5_8466371} and resource allocation  \cite{Zhang-infocom2019}, among others \cite{zhang2019deep}. 

\begin{figure}
  \centering
  \includegraphics[width=0.9\linewidth]{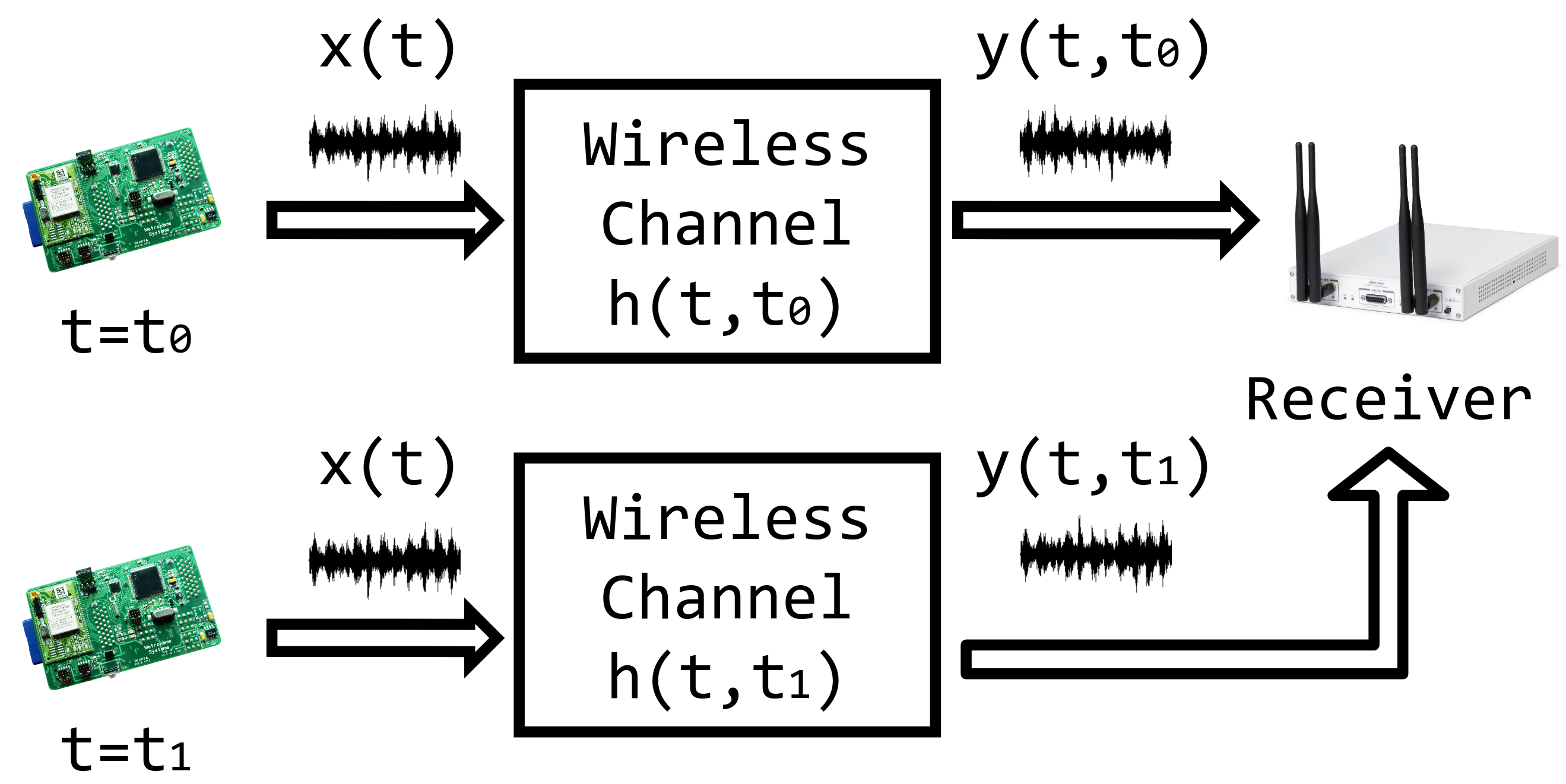}
   \caption{\label{fig:channel}The channel problem in wireless deep learning.\vspace{-0.6cm}}
\end{figure}

Although wireless deep learning has proven to be effective, recent research \cite{restuccia2019deepradioid,shawabka2020exposing} has demonstrated that the wireless channel can seriously compromise the effectiveness of the features learned by the classifier.  The key issue in wireless deep learning is that classifiers operating on waveforms are by definition \textit{time-varying} systems due to the ever-changing nature of the channel \cite{goldsmith2005wireless}. Figure \ref{fig:channel} shows a (simplified) representation of the channel issue in wireless deep learning. 

Addressing time variance  is one of modern machine learning's greatest issues \cite{Sugiyama:2012:MLN:2209761}, which makes the channel problem in wireless deep learning significantly challenging. One possible option is to periodically retrain the model; however,  deep learning networks usually require a significant time to be re-trained, even with modern GPUs \cite{chetlur2014cudnn}. Therefore, we cannot assume that \textit{the underlying deep learning model can be retrained in real time} before the channel has changed the features again. Another key reason to avoid the re-training of the model is the additional energetic burden imposed to the overall wireless infrastructure, which has to continously (i) collect new waveforms, (ii) fine-tune the model with the new data, and (iii) re-send the new weights to the wireless platform. Thus, this approach is highly inefficient -- and ultimately ineffective since it is hardly scalable to IoT networks.

 In other words, a waveform $x(t)$ transmitted at time $t_0$ and $t_1$ by the same device will necessarily be received with different levels of distortion by the receiver. Since the training process cannot include all the possible channel realizations and distributions, the features learned by the classifier cannot be entirely time-invariant \cite{Sankhe-infocom2019,sankhe2019noradio,5_8466371}. The seriousness of this problem has been recently exposed in the context of radio fingerprinting \cite{shawabka2020exposing}, where the authors have demonstrated that the accuracy of the classifier can drop down to almost zero when tested on I/Q samples collected in days different than the one when samples were collected for training.

To address this spinous issue, in our preliminary work \cite{restuccia2019deepradioid} we have introduced \emph{DeepRadioID}, a framework for channel-resilient optimization of deep learning-based radio fingerprinting algorithms. The core intuition was to impose a carefully-tailored modification of the waveform at the transmitter's side to ``restore'' the features compromised by the channel. To achieve this goal, the transmitter applies a digital finite input response filter (FIR) to the waveform before transmission, which slightly modifies its baseband signal to compensate for current channel conditions. The receiver is tasked with computing the FIR through mathematical optimization and then send it back to the transmitter as feedback, which will use the FIR to filter its own I/Q samples. We also demonstrated that an adversary cannot reuse the FIR  computed for another device to increase the probability to be identified as that device.

This paper builds upon and extends our work \cite{restuccia2019deepradioid} by making the following novel contributions:\smallskip

$\bullet$ We propose \emph{DeepFIR}, a real-time channel- and adversary-resilient optimization system to improve the accuracy of a \emph{generic} wireless deep learning model.  We postulate the \textit{Waveform Optimization Problem} (WOP) to find the optimal FIR, and derive a novel algorithm based on the \textit{Nonlinear Conjugate Gradient} (NCG) method  to efficiently solve it. We show in Section \ref{sec:ber} that the FIR's action can be effectively compensated at the receiver's side through the discrete Fourier transform (DFT) of the received signal, thus causing a negligible throughput decrease (\textit{i.e.}, less than 0.2\% in our experiments). We further propose a data-driven approach implemented in Keras to compute the optimal FIR without the need of external optimization algorithms (Section \ref{sec:data_driven}); \smallskip\smallskip

$\bullet$ We extensively evaluate the performance of \emph{DeepFIR} on an experimental testbed made up of 20 bit-similar devices (\textit{i.e}., transmitting the same baseband signal through nominally-identical RF interfaces and antennas). Testbed results (Section \ref{sec:exp_res}) indicate that (i) an adversary trying to imitate a fingerprint by using the same FIR filter \textit{decreases} its fingerprinting accuracy by about 54\% on the average; (ii) DeepFIR \textit{increases} the accuracy by 35\% on the average; \smallskip

$\bullet$ To further evaluate the scalability of \emph{DeepFIR} and to experiment with different wireless technologies, deeper learning models and learning problems we also leverage (i) two datasets of IEEE 802.11a/g (WiFi) and Automatic Dependent Surveillance -- Broadcast (ADS-B) transmissions, each containing 500 devices\footnote{Due to contract obligations, we cannot release the datasets to the community. We hope this will change in the future.} (Section \ref{sec:dataset_res}); and (ii) on a widely-available modulation recognition model \cite{OShea-ieeejstsp2018} trained on the RadioML 2018.01A dataset (Section \ref{sec:modulation_res}), which includes 24 different analog and digital modulations with different levels of signal-to-noise ratio (SNR);\smallskip

$\bullet$ Results on the large-scale datasets show that \emph{DeepFIR} increases the accuracy by 50\% and by 58\% on the 500-device ADS-B and WiFi datasets, respectively, and by comparing with the state of the art \cite{12_vo2016fingerprinting}, \emph{DeepFIR} improves the accuracy by about 27\% on a reduced dataset of 100 WiFi devices. Finally, our data-driven \emph{DeepFIR} methodology improves the accuracy of the modulation classifier of 2x on the average. \vspace{-0.3cm}

\section{Related Work}\label{sec:rw}

The topic of physical-layer deep learning has received significant attention over the last few years \cite{restuccia2020physicallayer}. The vast majority of existing work has applied carefully-tailored feature extraction techniques at the physical layer to fingerprint wireless devices \cite{8_nguyen2011device,12_vo2016fingerprinting,16-Peng-ieeeiotj2018,17-Xie-ieeeiotj2018,18-Xing-ieeecomlet2018} or to classify/optimize physical-layer parameters \cite{2_o2017introduction,OShea-ieeejstsp2018,Restuccia-infocom2019}.

One of the first papers to address radio fingerprinting is due to Nguyen \emph{et al.}~\cite{8_nguyen2011device}, who use device-dependent channel-invariant radio-metrics and propose a non-parametric Bayesian method to detect the number of devices. Recently, Vo \emph{et al.} \cite{12_vo2016fingerprinting} proposed a series of algorithms with features based on frequency offsets, transients and the WiFi scrambling seed, and validated them with off-the-shelf WiFi cards in a non-controlled RF environment, achieving accuracy between 44 and 50\% on 93 devices. Peng \textit{et al.}~\cite{16-Peng-ieeeiotj2018} proposed fingerprinting algorithms for ZigBee devices  based on modulation-specific features such as differential constellation trace figure (DCTF), showing that their features achieve almost 95\% accuracy on a 54-radio testbed. The usage of supervised machine learning techniques, moreover, has received a lot of attention over the last years, particularly the problem of modulation recognition \cite{Bkassiny-ieeecommsurtut2013,Jiang-ieeewcomm2017,Chen-arxiv2017}. Similar to radio fingerprinting, feature-based learning constitutes the majority of existing work \cite{Wong-isspa2001,Xu-ieeetvt2010,Pawar-ieeetifs2011,Shi-ieeetcomm2012,Ghodeshar-icscn2015}.

The key issue with legacy machine learning techniques is its inherent feature extraction process, which involves the computation of waveform characteristics such as kurtosis, high-order cyclic moments, median,  average and so on. Feature extraction is also very application-specific in nature, which hinders flexibility and adaptation. Thus, its applicability for real-time wireless spectrum analysis presents substantial limitations. Recently, deep learning \cite{lecun2015deep} has been proposed as a viable solution \cite{Mao-ieeecomm2018}. Deep learning has been extensively used to address modulation recognition \cite{OShea-ieeejstsp2018,o2017introduction,wang2017deep,West-dyspan2017,Kulin-ieeeaccess2018,Karra-ieeedyspan2017}.  In \cite{OShea-ieeejstsp2018}, O'Shea \emph{et al.} proposed several modulation classifiers, while in \cite{Karra-ieeedyspan2017} Karra \emph{et al.} use deep neural networks to classify modulation class and order.

Portion of the above work does consider the issue of signal-to-noise (SNR) ratio and other waveform impairments when evaluating the model performance. However, it does not propose any systematic investigation of the role of the wireless channel in the overall performance. To the best of our knowledge, Shawabka \emph{et al.} \cite{shawabka2020exposing} was the first work to investigate the issue, by definitely establishing through extensive performance evaluation that the time-varying nature of the channel impacts significantly the classification accuracy. In our previous work \cite{restuccia2019deepradioid}, we have proposed \emph{DeepRadioID}, a framework to improve the channel resilience of deep learning-based radio fingerprinting algorithms through the application of carefully-optimized finite response input (FIR) filters. This paper substantially extends our prior work by proving that our FIR-based approach is applicable to address general deep learning models and not only radio fingerprinting models, by appling \emph{DeepFIR} to improve the performance of a modulation classifier proposed in \cite{OShea-ieeejstsp2018}. Moreover, we have designed and developed a Keras-based version of our system, which is able to run on GPU and thus reduce significantly the FIR computation time.

\vspace{-0.2cm}

\section{DeepFIR: An Overview} \label{sec:system}

We first discuss some key observations and motivations to motivate our design choices in Section \ref{sec:observations}. We then provide an in-depth description and a walk-through of the main steps involved in the fingerprinting process in Section \ref{sec:overview}.\vspace{-0.3cm}

\subsection{DeepFIR: Key Intuitions}\label{sec:observations}

The need to optimize the accuracy of wireless deep learning systems arises from the fact that the wireless channel is dynamic and almost unpredictable in nature. Moreover, hardware impairments such as I/Q imbalances, DC offset, phase noise, carrier/sampling offsets, and power amplifier distortions are extremely time-varying and dependent on a number of factors, such as local oscillator (LO) frequency \cite{sourour2004frequency} and current temperature of the RF circuitry \cite{johnson1966physical}. These considerations imply that we cannot assume waveforms as perfectly stationary -- hence the need for real-time optimization.

To address the complex nature of the problem, our first observation is that convolutional neural networks (CNNs) have shown to be prodigiously suited to recognize complex ``patterns'' in input data \cite{lecun2015deep} -- these patterns are, in our case, the imperfections in the radio hardware. However, a major challenge that still lingers is \textit{how do we optimize the CNN output for a given device without retraining the CNN itself.} To answer this question, we devise a new approach based on finite impulse response (FIR) filtering of the transmitter's baseband signal to ``restore'' the patterns that are disrupted by the current channel conditions -- thus making the signal ``more recognizable'' to the CNN. We use FIRs because of the following: (i) FIRs are very easy to implement in both hardware and software on almost any wireless device; (ii) the computation complexity of applying a FIR filter of length $m$ to a signal is $\mathcal{O}(m)$ -- thus it is a very efficient algorithm; and most importantly, (iii) its effect on the BER can be almost perfectly compensated at the receiver's side, as shown in Section \ref{sec:ber}.

However, this approach spurs another challenge, which is \textit{how to set the FIR taps in such a way that the classification accuracy for a given class is maximized}. Our intuition here is to find the FIR that modifies input $x$ so that the resulting $x^*$ signal maximizes the neuron activation correspondent to a given class, as shown in Section \ref{sec:gradient}. We are able to do this efficiently since the layers inside CNNs, although non-linear, are derivable, and thus we can compute the gradient of the output with respect to the FIR taps according to a given input. This way, we can design an optimization strategy that is fundamentally general-purpose in nature. \vspace{-0.4cm}

\subsection{DeepFIR: A Walk-Through}\label{sec:overview}

\begin{figure*}
  \centering
  \includegraphics[width=0.9\linewidth]{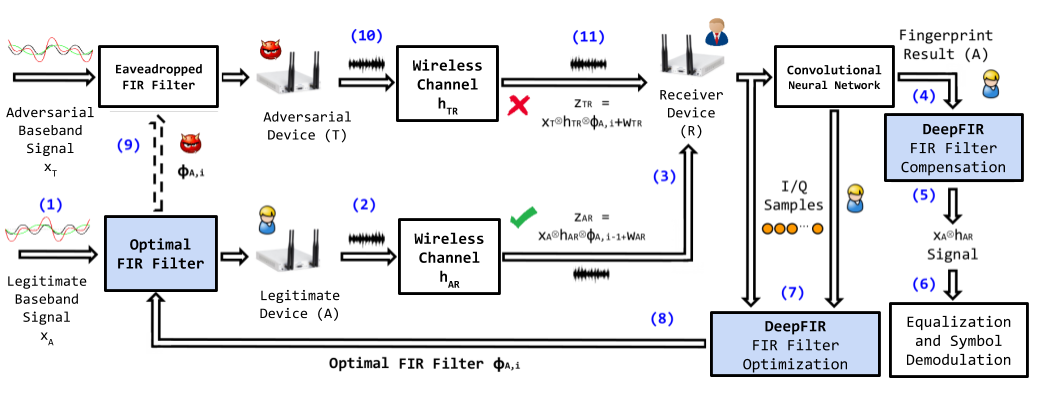}
   \caption{\label{fig:arch}A high-level overview of the \emph{DeepFIR} system, where we also illustrate an adversary (T) trying to impersonate a legitimate device (A) using an eavesdropped FIR filter. Since A's FIR filter has been tailored to match A's unique channel conditions, we show in Section \ref{sec:exp_res} that T does not fool the classifier by using A's filter.}\vspace{-0.5cm}
\end{figure*}

Figure \ref{fig:arch} provides a walk-through of the main building blocks of \textit{DeepFIR} and the main operations involved in the classification process. We highlight with a shade of blue the blocks that are added to the normal modulation/demodulation chain as part of \textit{DeepFIR}. The walk-through also shows how an adversary may try to camouflage its waveform of being of a different class (e.g., imitate the fingerpritint of another device). The detailed explanation of \emph{DeepFIR}'s main module will be given in Section \ref{sec:filt_design}.

The first step for a legitimate device ``A" that transmits a waveform to a receiver ``R'' is to filter its baseband signal with FIR $\bs \phi_{A,i-1}$ (step 1), which was obtained at the previous optimization step. FIR $\phi_{A, 0}$ is set to 1 (\textit{i.e.}, no filtering). The filtered signal is then sent to A's RF interface (step 2). By also accounting the effect of the wireless channel, ``R" will receive a baseband signal ${\bf z}_{AR} = {\bf x}_A \circledast \bs \phi_{A,i-1} \circledast \bf{h}_{AR} + {\bf w}_{AR}$, where ${\bf x}_A$ is the transmitted symbol sequence, $\bf{h}_{AR}$ and ${\bf w}_{AR}$ are the fading and noise introduced by the channel, respectively. The I/Q samples of ${\bf z}_{AR}$ are then fed to a CNN to classify the waveform (step 4). The fingerprinting result is then used to compensate the FIR filter $\bs \phi_{A,i-1}$ (step 5, discussed in Section \ref{sec:ber}), so that the resulting signal is then sent to the symbol demodulation logic to recover the application's data (step 6). The I/Q samples and the classification result are then fed to the \textit{DeepFIR} FIR Optimization module (step 7, presented in Section \ref{sec:filt_design}). The optimal FIR filter $\bs \phi_{A,i}$ is then sent back to A (step 8).

We now examine the case of adversarial action as follows. We assume that an adversarial device ``T" is capable of eavesdropping A's FIR $\bs \phi_{A,i}$. T's target here is to impersonate A by spoofing A's hardware fingerprint (step 9). After T's baseband signal is transmitted and goes through the wireless channel (step 10), the baseband signal received by R will be $\bf{z}_{TR} = \bf{x}_R \circledast \bs \phi_{A,i-1} \circledast \bf{h}_{TR} + \bf{w}_{TR}$, which is then fed to the CNN as input. However, we show in Section \ref{sec:exp_res} that, since A's FIR filter has been optimized for A's current wireless channel, T will not be able to utilize A's FIR filter to disguise its signal.

\section{DeepFIR FIR Optimization}\label{sec:filt_design}

In this section, we describe in details the \emph{DeepFIR} FIR Filter Optimization module. We first provide some background notions in Section \ref{sec:background}, followed by our FIR-based waveform modification approach in Section \ref{sec:firbasedmod}. We then introduce the Waveform Optimization Problem (WOP) in Section \ref{sec:gradient}.\vspace{-0.4cm}

\subsection{Background Notions and Definitions}\label{sec:background}
Let us define as \emph{input} a set of $N$ consecutive I/Q samples. Let us also define as \emph{slice} a set of $S$ inputs, and as batch a set of $B$ slices. Let us label the $D$ devices being classified with a label between 1 and $D$. We model the classifier as as a function $f : \mc X \rightarrow \mc Y$, where $\mc X \subseteq \mathbb{C}^{N}$ and $\mc Y \subseteq \mathbb{R}^{D}$ represent respectively the spaces of the classifier's input (\textit{i.e.}, an example) and output (\textit{i.e.}, a probability distribution over the set of $D$ devices).
Specifically, the output of the classifier can be represented as a vector $(f_1, f_2, ..., f_D)\in\mc Y$, where the $i$-th component denote the probability that the input fed to the CNN belongs to device $i$.

\textit{DeepFIR} relies on discrete finite impulse response filters (in short, FIRs) to achieve adaptive waveform modification. FIRs have the advantage that causal filters do not depend on future inputs, but only on past and present ones. Second, they can be represented as a weighted and finite term sum, which allows to \textit{accurately predict the output of the FIR for any given input}. More formally, a FIR is described by a finite sequence $\bs \phi$ of $M$ \textit{filter taps}, i.e., $\bs \phi=(\phi_1,\phi_2,\dots,\phi_M)$. 
For any input $\mf x \in \mc X$, the filtered $n$-th element $\hat{x}[n] \in \hat{\mf x}$ can be written as  
\begin{equation} \label{eq:FIR:general}
    \hat{x}[n] = \sum_{j=0}^{M-1} \phi_j x[n-j]
\end{equation}

Since the wireless channel operates in the complex domain by rotating and amplifying/attenuating the amplitude of the signal, \textit{we can manipulate the position in the complex plane of the transmitted I/Q symbols}. By using complex-valued filter taps, \textit{i.e.}, $\phi_k\in \mb C$ for all $k=0,1,\dots,M-1$, we can rewrite Eq. \eqref{eq:FIR:general} as follows:

\begin{align} 
\hat{x}[n] & = \sum_{k=0}^{M-1} (\phi^R_k +j\phi^I_k) (x^R[n-k]+jx^I[n-k]) \nonumber \\
& = \hat{x}^R[n] +  j \hat{x}^I[n]\label{eq:FIR:complex}
\end{align}
\noindent
where $x^R_k[n] = \mbox{Re}\{x_k[n]\}$, $x^I_k[n] = \mbox{Im}\{x_k[n]\}$, $\phi^R_k = \mbox{Re}\{\phi_k\}$ and $\phi^I_k = \mbox{Im}\{\phi_k\}$. Eq. \eqref{eq:FIR:complex} clearly shows that it is possible to manipulate the input sequence by filtering each I/Q sample. For example, to rotate all I/Q samples by $\theta=\pi/4$ radiants and halve their amplitude, we may set $\phi_1 = \frac{1}{2} \exp^{j\frac{\pi}{4}}$ and $\phi_k = 0$ for all $k>1$. 

\subsection{FIR-based Waveform Modification}\label{sec:firbasedmod}


Although channel equalization can effectively reduce the effect of channel distortions on the position of the received I/Q samples, the algorithms involved are generally not perfect and only \textit{partially} counteract phase and amplitude variations caused by the channel. For this reason, we must devise techniques to dynamically adapt to rapidly changing channel conditions (\textit{e.g.}, fast-fading/multi-path) and thus improve the  accuracy for a given class.

\begin{figure}[!h]
    \centering
    \includegraphics[width=\columnwidth]{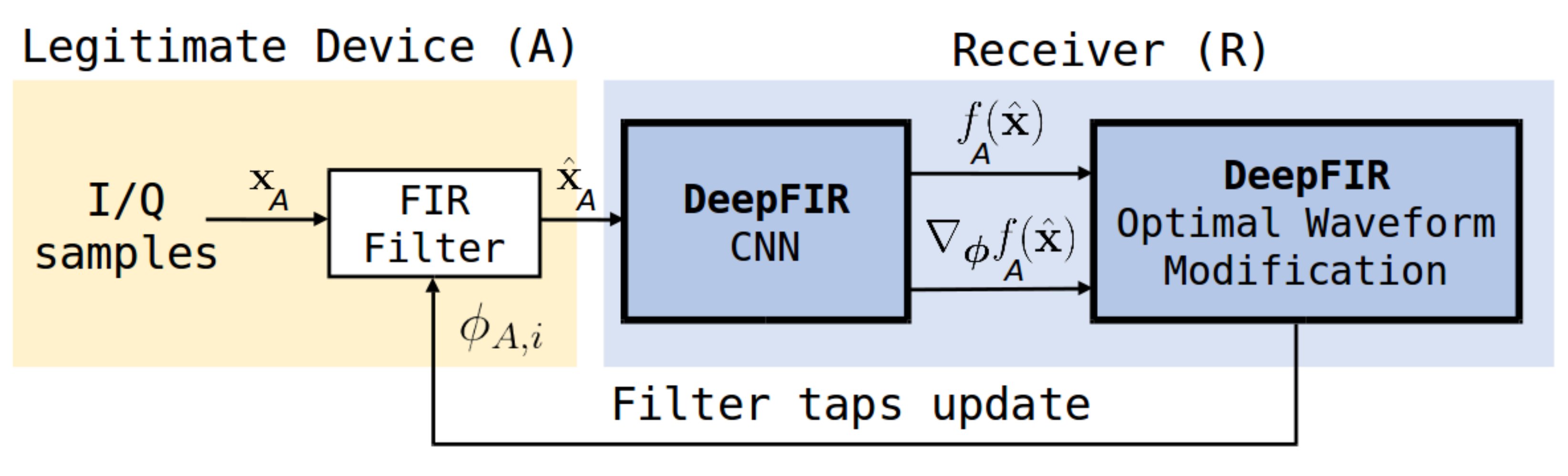}
    \caption{Waveform modification optimization loop.}
    \label{fig:DeepFIR}
\end{figure}

Figure \ref{fig:DeepFIR} shows a block diagram of the waveform modification optimization loop performed by \emph{DeepFIR}. Specifically, we add a FIR filter before the first CNN layer. This additional layer uses FIRs to manipulate the input example according to Eq.~\eqref{eq:FIR:complex}. The corresponding output sequence is then fed to the CNN. 

As shown in Figure \ref{fig:arch}, let $A$ be the target class for which we want to improve the detection accuracy of the CNN, and let $\bs \phi_{A,i}$ be the filter taps associated to the target class at the $i$-th optimization step. By using the filtering-based waveform modification on the input sequence $\bf x$, the output $f_A(\hat{\bf x}) \in\mc Y$ of the classifier with respect to the filtered sequence $\hat{\bf x}$ can be written as a function of the filter taps ${\bs \phi}_{A,i}$. Specifically, we have that 
\begin{equation} \label{eq:output}
    f_A(\hat{\bf x}) = f_A(\bf x, {\bs \phi}_{A,i})
\end{equation}

Eq.~\eqref{eq:output} clearly shows that the accuracy of the classifier depends on the actual FIR tap vector ${\bs \phi}_{A,i}$. Thus, we are interested in devising mechanisms to optimally manipulate ${\bs \phi}_{A,i}$ such that (i) the classification accuracy for the target class is maximized (Section \ref{sec:gradient}); and (ii) the waveform modification does not negatively impact the BER of data transmission activities (Section \ref{sec:ber}). To simplify the notation, henceforth we will remove the $i$ subscript.\vspace{-0.3cm}

\subsection{Waveform Optimization Problem (WOP)} \label{sec:gradient}

We can now formally define the objective of \textit{DeepFIR} as follows: (i) maximize the accuracy of the classifier for a specific target class $A$; and (ii) to guarantee that the resulting BER does not exceed a given maximum tolerable threshold $\mathrm{BER_{max}}$. Since we aim at achieving channel-resilient waveform modification, we need to compute a FIR parameter configuration ${\bs \phi}_A$ that can be used for multiple consecutive transmissions. To compute different ${\bs \phi}_A$ values for each single input $\bf x$ is inefficient in many cases, \textit{i.e.}, if applied to another input sequence $\bf x' \neq \bf x$, the FIR might decrease the accuracy of the classifier. 

To overcome the above problem, rather than maximizing the accuracy of the classifier on an input-by-input basis, we compute the the FIR ${\bs \phi}_A$ that maximizes the activation probability $f_A$ over a set of $S$ consecutive inputs, \textit{i.e.}, a slice.

The Waveform Optimization Problem (WOP) can be then defined as follows:
\begin{align}
\underset{\bs \phi}{\text{maximize}} & \hspace{0.2cm} \frac{1}{S} \sum_{s=1}^S f_A({\bf x}_s, \bs \phi) \label{prob:optimal:util} \tag{WOP} \\
    \text{subject to} & \hspace{0.1cm}  \mathrm{BER}_A(\mathrm{{\bf x}_s, \bs \phi)} \leq \mathrm{BER_{max}}, \hspace{0.2cm} \forall{s=1,2,\dots,S} \label{prob:optimal:constraint} \tag{C1}
\end{align}
\noindent
where the objective function represents the per-slice average activation probability for device $A$, ${\bf x}_s$ is the $s$-th input of the slice, and the $\mathrm{BER}_A(\cdot)$ represents the BER function corresponding to transmissions from target node $A$. The function $f_A(\mathbf{x}_s, \boldsymbol{\phi})$ represents the CNN, and thus it outputs the probability that the input I/Q samples $\mathbf{x}$ belong to device $A$. Thus,  we compute a FIR that maximizes the activation probability of the neuron associated to $A$. 

Problem \eqref{prob:optimal:util} is significantly challenging because (i) the function $f_A$ is CNN-specific and depends from a very high number of parameters (generally in the order of millions), it is highly non-linear and to the best of our knowledge, there are no mathematical closed-form expressions for such a function, even for relatively small CNNs; (ii) the maximum BER constraint \eqref{prob:optimal:constraint} depends from numerous device-specific parameters (\textit{e.g.}, modulation, coding, transmission power and SNR) and it is generally non-linear.

Notwithstanding the above challenges, and as we will discuss in detail in Section \ref{sec:ber}, the impact of the waveform modification procedure on the BER of communications among the receiver and the target device $A$ is negligible. Indeed, as shown in Figure \ref{fig:arch}, \textit{DeepFIR} embeds a FIR Filter Compensation module that uses peculiar features of FIR filters, \textit{e.g.}, their Fourier transform, to successfully reconstruct the original transmitted unfiltered sequence of I/Q symbols. This compensation procedure effectively removes any coupling between waveform modification procedures and BER, \textit{i.e.}, $\mathrm{BER}_A(\mathrm{\bf x, \bs \phi)} \approx \mathrm{BER}_A(\mathrm{\bf x})$.
Accordingly, it is possible to relax Constraint \eqref{prob:optimal:constraint} by removing it from the optimization problem \eqref{prob:optimal:util}.

The relaxed WOP can be formulated as  
\begin{align}
\underset{\bs \phi}{\text{maximize}} & \hspace{0.2cm} \sum_{s=1}^S f_A({\bf x}_s, \bs \phi) \label{prob:optimal:util:2} \tag{RWOP}
\end{align}
\noindent
where we have also omitted the constant term $1/S$.

\subsubsection{Solving the RWOP}

As already mentioned, $f_A$ is non-linear and generally does not possess any useful property in terms of monotonicity, concavity and existence of a global maximizer. 
However, for any input $\bf x$ of the slice, by using \textit{back-propagation} and the chain rule of derivatives it is possible to let the CNN compute the gradient $\nabla_{\hat{\bf x}}f_A(\hat{\bf x})$ of the classification function $f_A$ with respect to the filtered input sequence $\hat{\bf x}$. It is worth noting that $\nabla_{\hat{\bf x}}f_A(\hat{\bf x})$ shows how different input sequences affect the accuracy of the classification function. Nevertheless, we are interested in evaluating the gradients $\nabla_{\bs \phi}f(\hat{\bf x})$ to predict how the accuracy of the classifier varies as a function of the FIR filtering function.
Hence, we need to extend back-propagation to the waveform modification block. 

From Eq. \eqref{eq:FIR:complex}, $\hat{\bf x}$ is a function of $\bs \phi$, thus the gradient of $f_A$ with respect to the filter taps $\bs \phi$ can be computed as
\begin{equation} \label{eq:chain:vector}
    \nabla_{\bs \phi}f_A(\hat{\bf x}) = J_{f_A}(\bs \phi)^\top \cdot \nabla_{\hat{\bf x}}f_A(\hat{\bf x})
\end{equation}
\noindent
where $J_{f_A}(\bs \phi)$ is the Jacobian matrix of $f_A(\bf x, \bs \phi)$ with respect to $\bs \phi$, $\top$ is the transposition operator, and $\cdot$ stands for matrix dot product. 

From Eq. \eqref{eq:chain:vector} and Eq. \eqref{eq:FIR:complex}, each element in $\nabla_{\bs \phi}f_A(\hat{\bf x})$ can be written as
\begin{align} \label{eq:chain:partial}
    \frac{\partial f_A(\bf x, \bs \phi) }{\partial \phi^Z_k} & = \sum_{n = 1}^N \left( \frac{\partial f_A(\bf x, \bs \phi)}{\partial \hat{x}^R[n]} \frac{\partial \hat{x}^R[n] }{\partial \phi^Z_k} + \frac{\partial f_A(\bf x, \bs \phi)}{\partial \hat{x}^I[n]} \frac{\partial \hat{x}^I[n] }{\partial \phi^Z_k}\right)
\end{align}
\noindent
where $k=0,1,\dots,M-1$, $N$ is the length of the input sequence and $Z\in\{R,I\}$.

By using Eq. \eqref{eq:FIR:complex}, $\frac{\partial \hat{x}^R[n] }{\partial \phi^Z_k}$ and $\frac{\partial \hat{x}^I[n] }{\partial \phi^Z_k}$ in Eq. \eqref{eq:chain:partial} are computed as follows:
\begin{align}
    \frac{\partial \hat{x}^R[n] }{\partial \phi^R_k} & = \frac{\partial \hat{x}^I[n] }{\partial \phi^I_k} = x^R[M-1+n-k] \label{eq:chain:partial:1} \\
    \frac{\partial \hat{x}^I[n] }{\partial \phi^R_k} & = - \frac{\partial \hat{x}^R[n] }{\partial \phi^I_k} = x^I[M-1+n-k] \label{eq:chain:partial:2}
\end{align}

The above analysis shows that the relationship between the waveform modification and classification processes can be described by a set of gradients. Most importantly, they can be used to devise effective optimization algorithms that solve Problem \eqref{prob:optimal:util:2}.

In Section \ref{sec:ncg}, we design an algorithm to solve Problem \eqref{prob:optimal:util:2} and compute the optimal FIR filter parameters $\bs \phi$ by using the Nonlinear Conjugate Gradient (NCG) method and the gradients computed in Eq. \eqref{eq:chain:partial:1} and  Eq. \eqref{eq:chain:partial:2}. While our simulation results have shown that NCG is more accurate than other gradient-based optimization algorithms (\textit{e.g.}, gradient descent algorithms), we remark that \textit{DeepFIR} is independent of the actual algorithm used to compute $\bs \phi$, and other approaches can be used to solve Problem \eqref{prob:optimal:util:2}.

\subsubsection{Filter taps computation through NCG} \label{sec:ncg}

As shown in Figure \ref{fig:arch} and discussed in Section \ref{sec:system}, \textit{DeepFIR} iteratively adapts to channel fluctuations by periodically updating the filter taps associated to any given target device $A$. 
For the sake of generality, we refer to this periodic update as an \textit{optimization epoch}, and a new epoch is started as soon as one or more \textit{triggering events} are detected by \textit{DeepFIR}. Triggering events can be either cyclic, \textit{e.g.}, timer timeout, or occasional, \textit{e.g.}, the accuracy for a target devices falls below a minimum desired threshold.   

For each epoch $i$, let $t=1,2,\dots,T$ denote the iteration counter of the optimization algorithm. At each iteration $t$ of the algorithm, the filter taps are updated according to the following iterative rule: ${\bs \phi}^{(t)} = {\bs \phi}^{(t-1)} + {\bs \alpha^{(t)}} {\bf p^{(t)}}$, where ${\bf p^{(t)}}$ and ${\bs \alpha^{(t)}}$ represent the search direction and update step of the algorithm, respectively. To put it simple, ${\bf p^{(t)}}$ gives us information on the direction to be explored, while ${\bs \alpha^{(t)}}$ tells us how large the exploration step taken in that direction should be. More in detail, the two terms are computed as follows:
\begin{align}
    {\bf p^{(t)}} &= \sum_{s=1}^S \left( \nabla_{{\bs \phi}} f_A({\bf x}_s,{\bs \phi}^{(t-1)}) \right) + {\bs \beta}^{(t)} {\bf p^{(t-1)}} \label{eq:CGM:direction} \\
    {\bs \alpha}^{(t)} & = \argmax_{\bs \alpha} \sum_{s=1}^S f_A({\bf x}_s,{\bs \phi}^{(t-1)} + {\bs \alpha} {\bf p^{(t)}}) \label{eq:CGM:step}
\end{align}
\noindent
where gradients derive from Eq. \eqref{eq:chain:partial:1} and Eq. \eqref{eq:chain:partial:2}, $\beta^{(1)} = 0$ and $p^{(0)} = 0$.
The parameter ${\bs \beta}^{(t)}$ is defined as 
\begin{equation}
        {\bs \beta}^{(t)} = \frac{|| \sum_{s=1}^S \left( \nabla_{{\bs \phi}} f_A({\bf x}_s,{\bs \phi}^{(t-1)}) \right) ||^2_2}{|| \sum_{s=1}^S \left( \nabla_{{\bs \phi}} f_A({\bf x}_s,{\bs \phi}^{(t-2)}) \right) ||^2_2} \label{eq:CGM:beta}
\end{equation}
\noindent
and is generally referred to as the \textit{conjugate gradient} (update) parameter used in NCG methods to improve the space exploration process by speeding up the convergence of the algorithm \cite{hestenes1952methods}. 

Interestingly enough, ${\bf p^{(1)}} = \sum_{s=1}^S \left( \nabla_{{\bs \phi}} f_A({\bf x}_s,{\bs \phi}^{(0)}) \right)$, meaning that when $t=1$ NCG is a classic gradient descent. Also, ${\bs \alpha}^{(t)}$ in Eq. \eqref{eq:CGM:step} is computed through line-search algorithms. While both exact and approximated line search algorithms can be considered, there are few aspects that need to be considered when implementing Eq. \eqref{eq:CGM:step}. Indeed, since the function $f_A$ is highly non-linear and has no closed-form representation, to compute Eq. \eqref{eq:CGM:step} requires the continuous evaluation of $\sum_{s=1}^S f_A({\bf x}_s,{\bs \phi}^{(t-1)} + {\bs \alpha} {\bf p^{(t)}})$ and its first and second order derivatives. For this reason, it might be computationally expensive to run exact line search algorithms on $f_A$, and approximated line search algorithms are to be preferred. For example, to speed-up the computation of ${\bs \alpha}^{(t)}$, we can consider a secant method where the second derivatives of $f_A$ are approximated by using the first order derivatives computed in Eq. \eqref{eq:chain:partial:1} and Eq. \eqref{eq:chain:partial:2}.

\section{DeepFIR FIR Compensation}\label{sec:ber}

Although the waveform filtering process is beneficial to the classification process as we can optimally modify the waveform generated by a given target device, it may negatively affect the quality of transmitted data. Furthermore, the expression of the BER in Constraint \eqref{prob:optimal:constraint} is non-linear and possesses exact and/or approximated closed-form representations only in a limited number of cases (\textit{e.g.}, fading channels with known distributions and low-order modulations). To overcome the above issues, we observe that our waveform modification relies on Eq.~\eqref{eq:FIR:general}, which clearly represents a discrete convolution between the input sequence $\bf x$ and the filter taps $\bs \phi$. For the sake of illustration, in the following we use the familiar model
\begin{equation}
    z[n] = ({\bf h} \circledast {\bf \hat{x}})[n] + w[n]
\end{equation}
\noindent
where each received I/Q symbol $z[n]$ is written as the sum of a noise term $w[n]$ (typically Additive white Gaussian noise) and the $n$-th element of the discrete convolution ${\bf h} \circledast {\bf \hat{x}}$ between the channel $h$ and the transmitted filtered sequence ${\bf \hat{x}}$. 
From Eq.~\eqref{eq:FIR:general}, we also have that $\hat{{\bf x}} = {\bf x} \circledast \phi$. 
Thus, the discrete Fourier transform (DFT) of ${\bf z}$ and ${\bf \hat{x}}$ can be written as follows:\vspace{-0.2cm}

\begin{align} \label{eq:DFT}
    Z(\omega) & = H(\omega) \hat{X}(\omega) + W(\omega) \nonumber \\
         & = H(\omega) X(\omega) \Phi(\omega) + W(\omega)
\end{align}
\begin{equation} \label{eq:DFT:X}
    X(\omega) = \frac{Z(\omega) - W(\omega)}{H(\omega) \Phi(\omega)}
\end{equation}
\noindent
where we have used capital letters to indicate DFTs.

Eq.~\eqref{eq:DFT:X} shows that to reconstruct the original unfiltered I/Q sequence is possible by computing the inverse DFT of each component of the received signal. Furthermore, FIR compensation implies $\mathrm{BER}_A(\mathrm{\bf x, \bs \phi)} = \mathrm{BER}_A(\mathrm{\bf x})$. Since, the optimization variable of Problem \eqref{prob:optimal:util} is $\bs \phi$, Constraint \eqref{prob:optimal:constraint} does not depend on $\bs \phi$ anymore, and thus it can be removed from Problem \eqref{prob:optimal:util}. Finally, note that the above FIR compensation method is independent of the underlying modulation and coding scheme. This means that FIR compensation is a general approach, and it can be successfully used to tackle Constraint \eqref{prob:optimal:constraint}.

Despite the above properties, one might argue that in general it is not possible to compute a perfect estimation of $W(\omega)$ and $H(\omega)$. However, modern wireless networks embed estimation mechanisms that are almost always able to compute fairly accurate estimations $\tilde{H}(\omega)$ and $\tilde{N}(\omega)$, \textit{e.g.}, through training sequences and pilots \cite{sourour2004frequency}, and the effect of the FIR filter can thus be compensated to a significant extent. To validate this crucial assumption, we ran a number of experiments on our experimental testbed to evaluate the impact of \emph{DeepFIR}'s FIR filtering on the packet error rate (PER) and throughput ($\theta$) of a wireless transmission. 

\begin{figure}[!h]
    \centering
    \includegraphics[width=0.9\columnwidth]{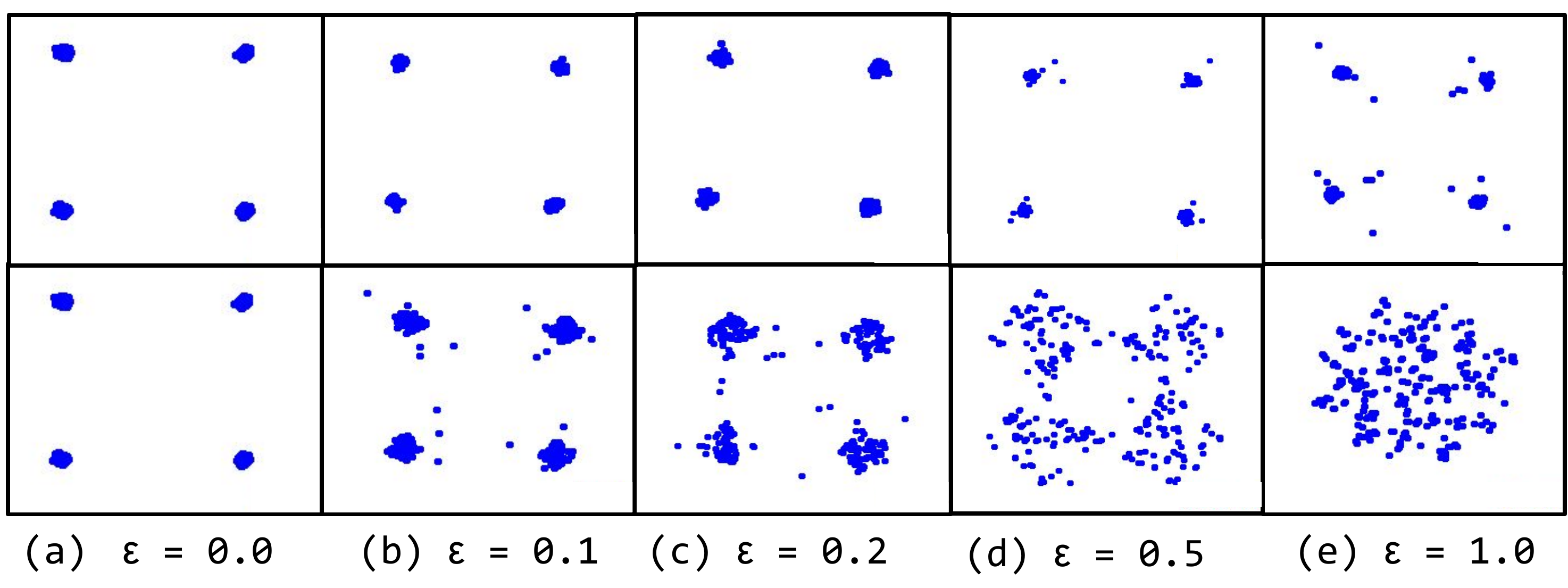}
    \caption{The effect of FIR filtering at the receiver's side for different values of $\epsilon$. Top and bottom sides show respectively the received constellations with/without FIR compensation.}
    \label{fig:FIR_effect}
\end{figure}

\begin{figure}[!h]
    \centering
    \includegraphics[width=0.43\columnwidth,angle=-90]{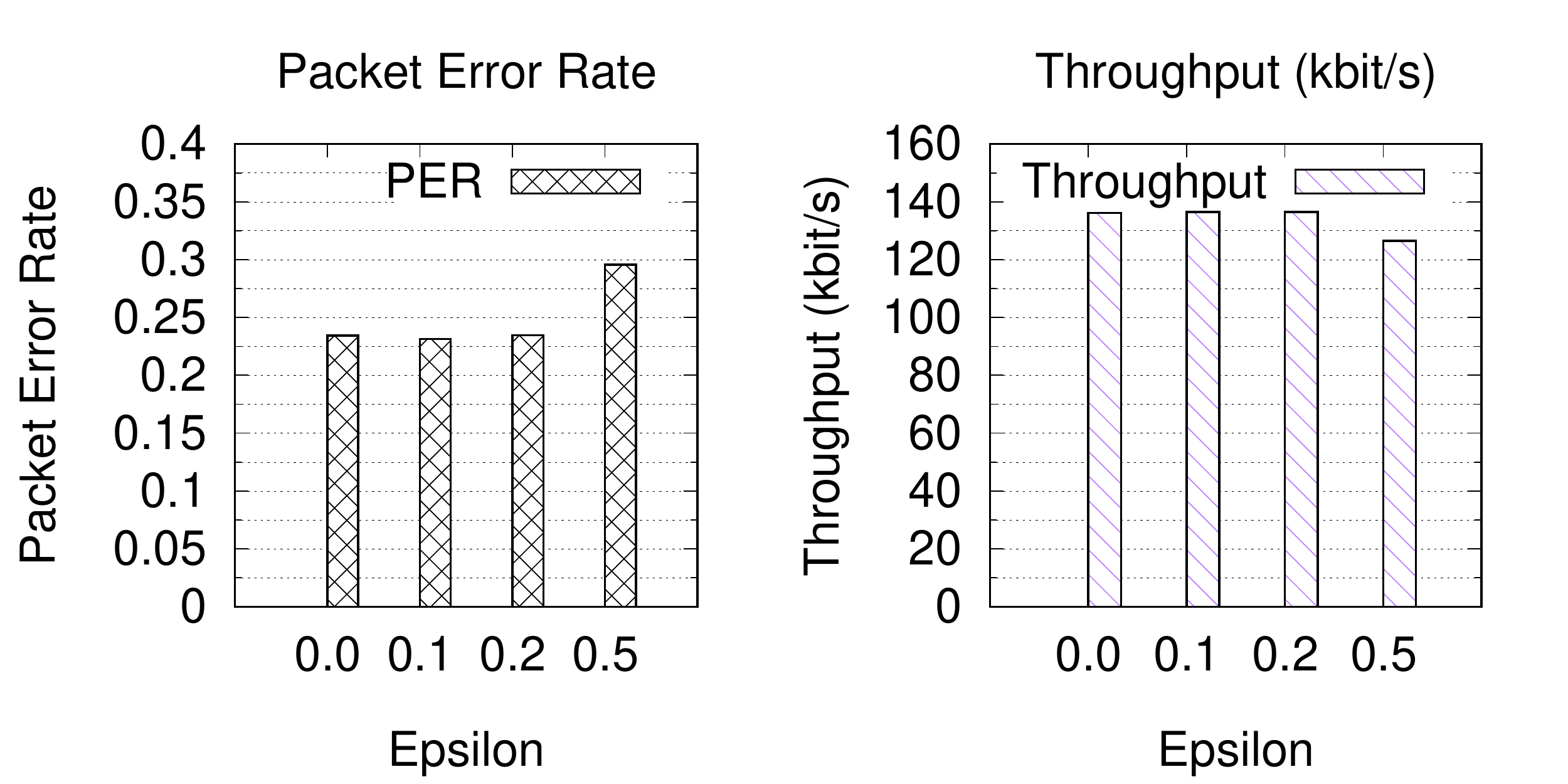}
    \caption{Packet error rate (PER) and Throughput ($\theta$) as a function of $\epsilon$.\vspace{-0.5cm}}
    \label{fig:PER_Throughput}
\end{figure}

Figure \ref{fig:FIR_effect} shows the received constellations of a QPSK-modulated WiFi transmission where the payload I/Q samples are multiplied in the frequency domain (\textit{i.e.}, FIR filtering) with a random I/Q tap with $ I \in [1 -\epsilon, 1 + \epsilon]$ and $Q \in [0 -\epsilon, 0 + \epsilon])$. The $\epsilon$ parameter represents the relative magnitude of the filter with respect to no filtering, \textit{i.e.}, $\epsilon = 0$. The filtering is then compensated at the receiver's side by using Eq.~\eqref{eq:DFT:X}. 

We notice than some noise is indeed introduced by the FIR filtering irrespective of our compensation. However, Figure \ref{fig:PER_Throughput} shows the PER and $\theta$ as a function of $\epsilon$, which respectively increase and decrease of about 6\% and 0.5~kbit/s in the worst case of $\epsilon = 0.5$. However, according to our experiments in Section \ref{sec:exp_res}, the $\epsilon$ value is typically below 0.2, meaning a PER increase $<$1\% and a $\theta$ loss $<$0.2 kbit/s (0.2\%). 


\section{Data-driven FIR Optimization}\label{sec:data_driven}

Although traditional optimization techniques can indeed be leveraged to compute FIR taps for each individual class, these techniques algorithms usually suffer from scalability issues , primarily due to the complexity of computing an optimal solution. Although this might not be an issue when the number of classes and/or the size of the dataset are small, the same does not hold anymore when both the problem and the inputs grow in size. Also, while trained models leverage well-established machine learning software libraries (such as Keras, TensorFlow and PyTorch) that take advantage of GPU processing to speed-up computation, the majority of optimization numerical solvers (\textit{e.g.}, Matlab, CPLEX) run on CPUs and do not directly interface with neural network models. A direct consequence of this latter aspect is that the numerical solver and the training model run on separate environments, which inevitably introduces computation delays especially when computing gradients. 
To overcome the above limitations, in this section we present a novel data-driven approach to compute FIR taps for each class individually. \textbf{The code repository is available for download at \url{https://github.com/frestuc/DeepFIR}.}

Equation \eqref{eq:FIR:complex} shows that FIRs can be modeled as a discrete convolution between FIR taps and the input data. Thus, the same operation can be implemented as an \textit{FIR Layer}, whose weights represent FIR taps and whose output follows \eqref{eq:FIR:complex}. Since each class requires an individual FIR tap configuration, we can generate $D$ distinct FIR Layers whose weights, \textit{i.e.}, the FIR taps, are trained individually on a per-class basis.

\begin{figure}[!h]
    \centering
    \includegraphics[width=0.9\columnwidth]{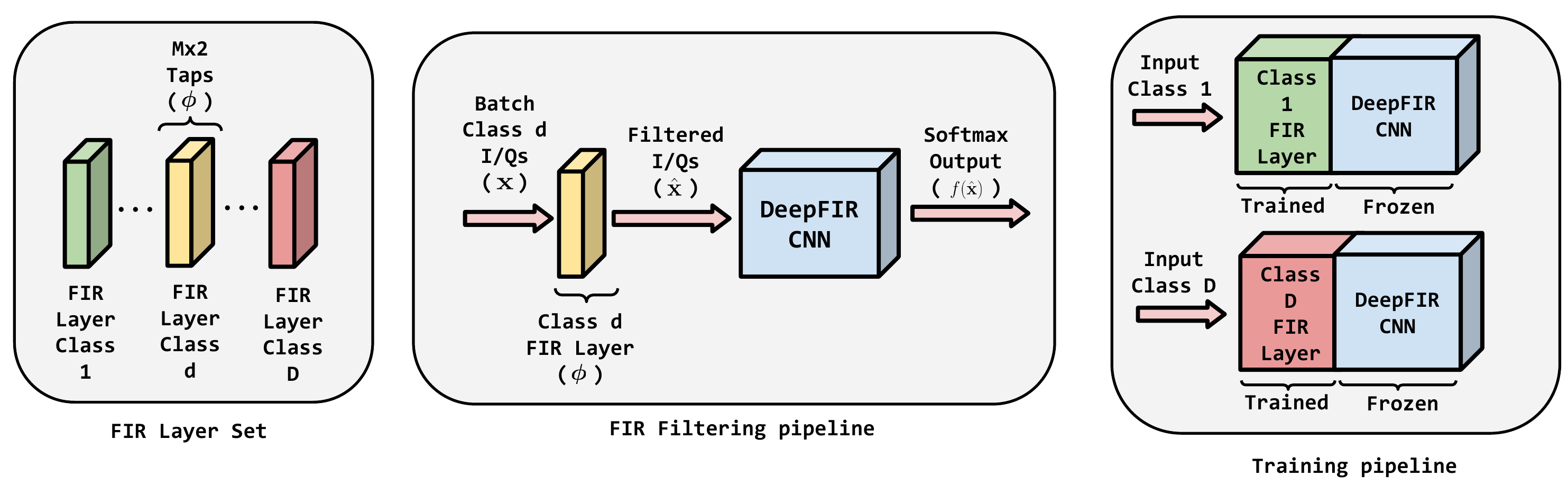}
    \caption{FIR layer structure, filtering and training pipeline for data-driven FIR computation.\vspace{-0.3cm}}
    \label{fig:FIR_layer}
\end{figure}

Figure \ref{fig:FIR_layer} shows a simplified block scheme summarizing our data-driven methodology to compute FIR filters. We generate $D$ FIR Layers, each consisting of $M\times2$ weights (\emph{i.e.}, real and imaginary part of the $M$ FIR taps), \textit{i.e.}, $\boldsymbol{\phi}=(\phi^R_k,\phi^I_k)_{k=0,\dots,M-1}$. We also load DeepFIR CNN pre-trained weights. As shown in Figure \ref{fig:FIR_layer} (right), only FIR Layer weights $\boldsymbol{\phi}$ are trained, while DeepFIR CNN model's weights are frozen. We can think of each FIR Layer as additional convolutional layer that is attached to the pre-trained DeepFIR CNN model.

Each FIR layer is fed with I/Q samples belonging to a specific class and it is trained such that filtered I/Q samples generated by the corresponding FIR layer maximize the accuracy of the pre-trained frozen model. We remark that \textit{we do not change the loss function $L$ of the pre-trained DeepFIR CNN model}. Instead, and as shown in Figure \ref{fig:FIR_layer}, we train FIR Layer $\boldsymbol{\phi}$, use them to manipulate the input $\mathbf{x}$ and fed the filtered input $\mathbf{\hat{x}}$ (computed as in \eqref{eq:FIR:complex}) to DeepFIR CNN. 
Each class's FIR Layer weights $\boldsymbol{\phi}$ are trained by minimizing the cross-entropy loss function $L$:

\begin{equation}
    L = \sum_{d = 1}^D y_d(\mathbf{\hat{x}}) \log(f_d(\mathbf{\hat{x}})) = \sum_{d = 1}^D y_d(\mathbf{x}) \log(f_d(\mathbf{x},\boldsymbol{\phi}))
\end{equation}
\noindent
where $\mathbf{x}$ represents the unfiltered input (\textit{i.e.}, a batch of $S$ slices as discussed in Section~\ref{sec:gradient}), $\mathbf{\hat{x}})$ is the filtered input sequence, $y_d(\mathbf{x})$ is a binary indicator such that $y_d(\mathbf{x})=1$ if the input $\mathbf{x}$ belongs to class $d$, and $f_d$ is the soft-max activation of the last layer of the pre-trained DeepFIR CNN model. 

Our approach, although it does not necessarily guarantee a global optimum,  enjoys several useful properties: (i) FIR taps are computed on GPUs via machine learning libraries; and (ii) there is no need to interface the trained model with external numeric solvers. Together, these two features makes it possible not only to speed-up FIR computation, but also to train FIR layers over extremely large datasets (such as those presented in Section \ref{sec:dataset_res}.

To avoid excessive distortion of input signals, FIR Layer weights are initialized such that the first weight value $\phi^R_0$ is set to 1 and the remaining taps are set to 0. This initialization in essence represents an identity vector, which returns unchanged input values. To allow for learning and different gradient updates, the 0’s are masked with small Gaussian noise. This acts as a form of "personalization" of filters: a device-specific filter is designed to enhance its detectability. This filter is presumed to be attached to the device, and hence is used both in the training and testing phases. This enables the filters to learn channel effects in the scope of individual devices.\vspace{-0.3cm}

\section{Radio Testbed Results}\label{sec:exp_res}




\subsection{Wireless Testbed and Data Collection}\label{sec:wireless_test}

Our experimental testbed is composed by twenty software-defined USRP radios acting as transmitters and one USRP acting as receiver. Each USRP has been equipped with a CBX 1200-6000 MHz daughterboard with 40 MHz instantaneous bandwidth \cite{CBX} and one VERT2450 antennas \cite{Vert2450}. Therefore, the RF components of each USRP are nominally-identical. Furthermore, each USRP device sends the same baseband signal, \textit{i.e.}, an IEEE 802.11a/g (WiFi) frame repeated over and over again, to make sure that the deep learning model is learning the hardware impairments and not data patterns. 

\begin{figure}[!h]
    \centering
    \includegraphics[width=0.9\columnwidth]{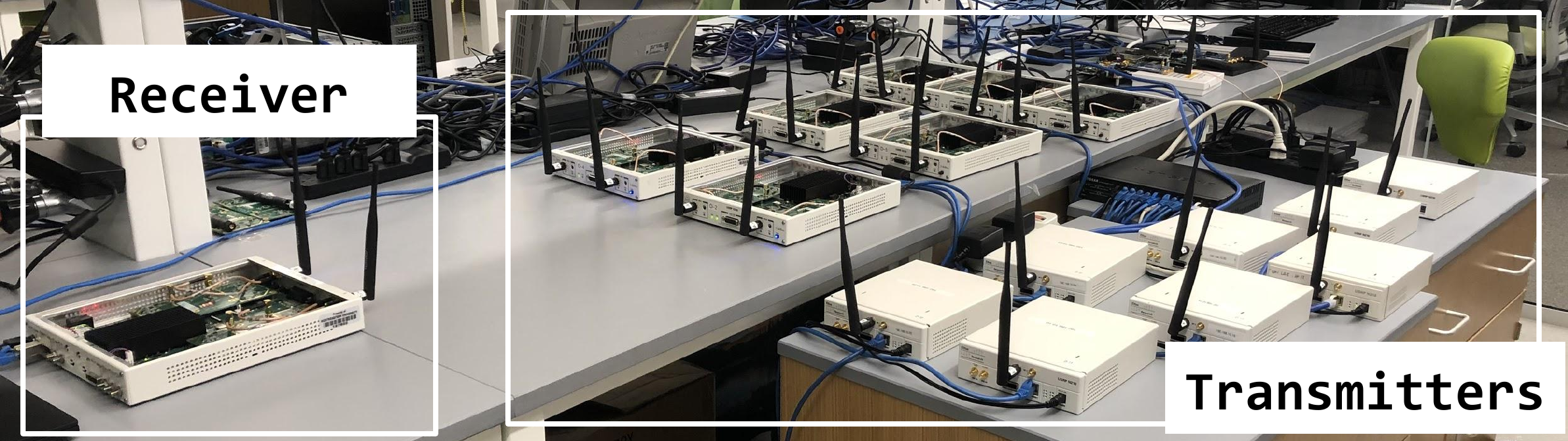}
    \caption{DeepFIR Experimental Testbed.\vspace{-0.7cm}}
    \label{fig:testbed}
\end{figure}

The baseband signal is generated through Gnuradio and then streamed to the selected SDR for over-the-air wireless transmission. The receiver SDR samples the incoming signals at $10\ \textrm{MS/s}$ sampling rate at center frequency of $2.432\  \textrm{GHz}$. The collected baseband signal is then channel-equalized using IEEE 802.11 pilots and training sequences \cite{sourour2004frequency}. Next, the payload I/Q samples are extracted and partitioned into a \textit{sample}. In our experiments, we fix the sample length to $ 48 \cdot 6 = 288$ I/Q values, corresponding to 6 OFDM symbols containing 48 payload I/Q values. Each of these samples are then used for training and classification. \vspace{-0.5cm}

\subsection{Deep Learning Architecture}\label{sec:deep_learn_arch}

\begin{figure}[!h]
    \centering
    \includegraphics[width=0.8\columnwidth]{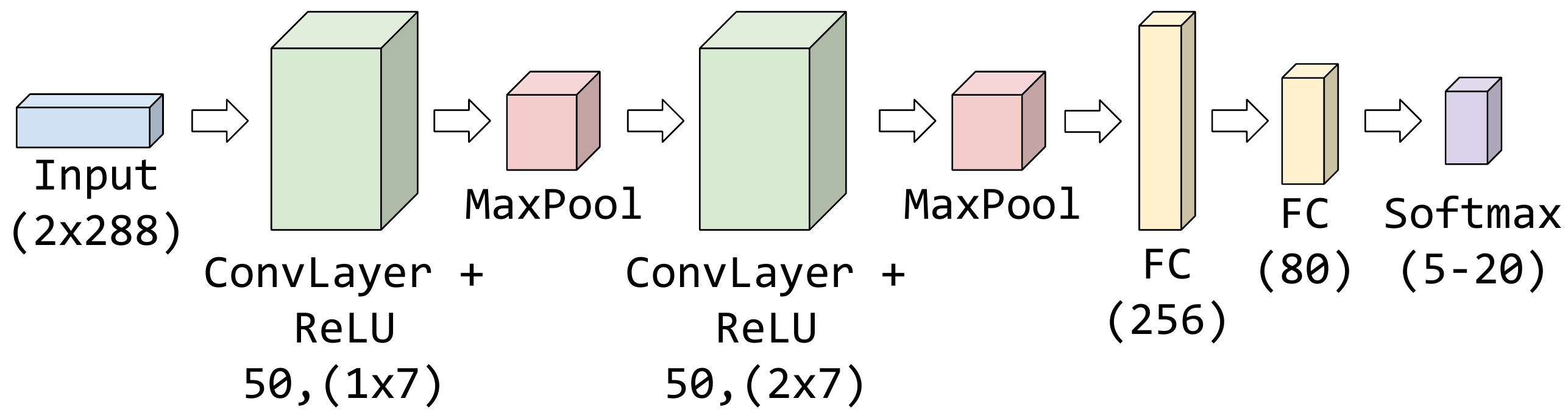}
    \caption{CNN used in our radio testbed experiments.\vspace{-0.3cm}}
    \label{fig:cnn_testbed}
\end{figure}

We use the CNN architecture reported in \cite{5_8466371} and depicted in Figure \ref{fig:cnn_testbed}. Specifically, each I/Q input sequence is represented as a two-dimensional real-valued tensor of size $2 \times 288$. This is then fed to the first convolutional  layer (ConvLayer), which consists of 50 filters  each of size $1 \times 7$. Each filter learns a 7-sample variation in time over the I or Q dimension separately, to generate 50 distinct feature maps over the complete input sample. Similarly, the second ConvLayer has 50 filters each of size $2 \times 7$.  Each ConvLayer is followed by a Rectified Linear Unit (ReLU) activation and a maximum pooling (MaxPool) layer with filters of size $2 \times 2$ and stride $1$, to perform a pre-determined non-linear transformation on each element of the convolved output.

We use an $\ell_2$ regularization parameter $\lambda=0.0001$. The weights of the network are trained using Adam optimizer with a learning rate of $l = 0.0001$. We minimize the prediction error through back-propagation, using categorical cross-entropy as a loss function computed on the classifier output. To train our CNN, we constructed a dataset by performing 10 different transmissions of length 5 minutes for each of the 20 devices, each transmission spaced approximately 5 minutes in time. 

\subsubsection{Performance Metrics}\label{sec:train_test}

Hereafter, we will use the following two metrics to assess the performance of our learning system:

\begin{enumerate}
    \item \emph{Per-Slice Accuracy (PSA).} As in Section \ref{sec:gradient}, a \textit{slice} is a set of $S$ consecutive input. The PSA is thus defined as the average CNN fingerprinting accuracy on the $S$-sample slice. Since the FIR filter optimized by \emph{DeepFIR} is computed on one slice, the PSA measures how much \emph{DeepFIR} is able to increase the short-term accuracy of the CNN.
    \item \emph{Per-Batch Accuracy (PBA).} We define as \textit{batch} as a set of $B$ consecutive slices. The PBA is thus defined as the average CNN fingerprinting accuracy on the $B$-slice batch. The PBA measures the impact of the optimal FIR filter on the long-term accuracy of the CNN.
\end{enumerate}

In the following experiments, we use live-collected data collected 7 days after the data used for training the dataset was collected. To allow experiments' repeatability, we record a transmission from a given device for about one minute. Then, we select the first slice from the recording and compute the PSA with no FIR optimization. Next, we perform FIR optimization on the slice, re-filter each sample in the slice, and compute the PSA after FIR filtering. To compute the PBA, we apply the same FIR filter to the  $B \cdot N -1$ slices collected after the first one and then compute the average PSA on the $B$-slice batch.\vspace{-0.5cm}

\subsection{Results}\label{sec:opt_res}

\begin{figure}[!h]
    \centering
    \includegraphics[width=0.4\columnwidth,angle=-90]{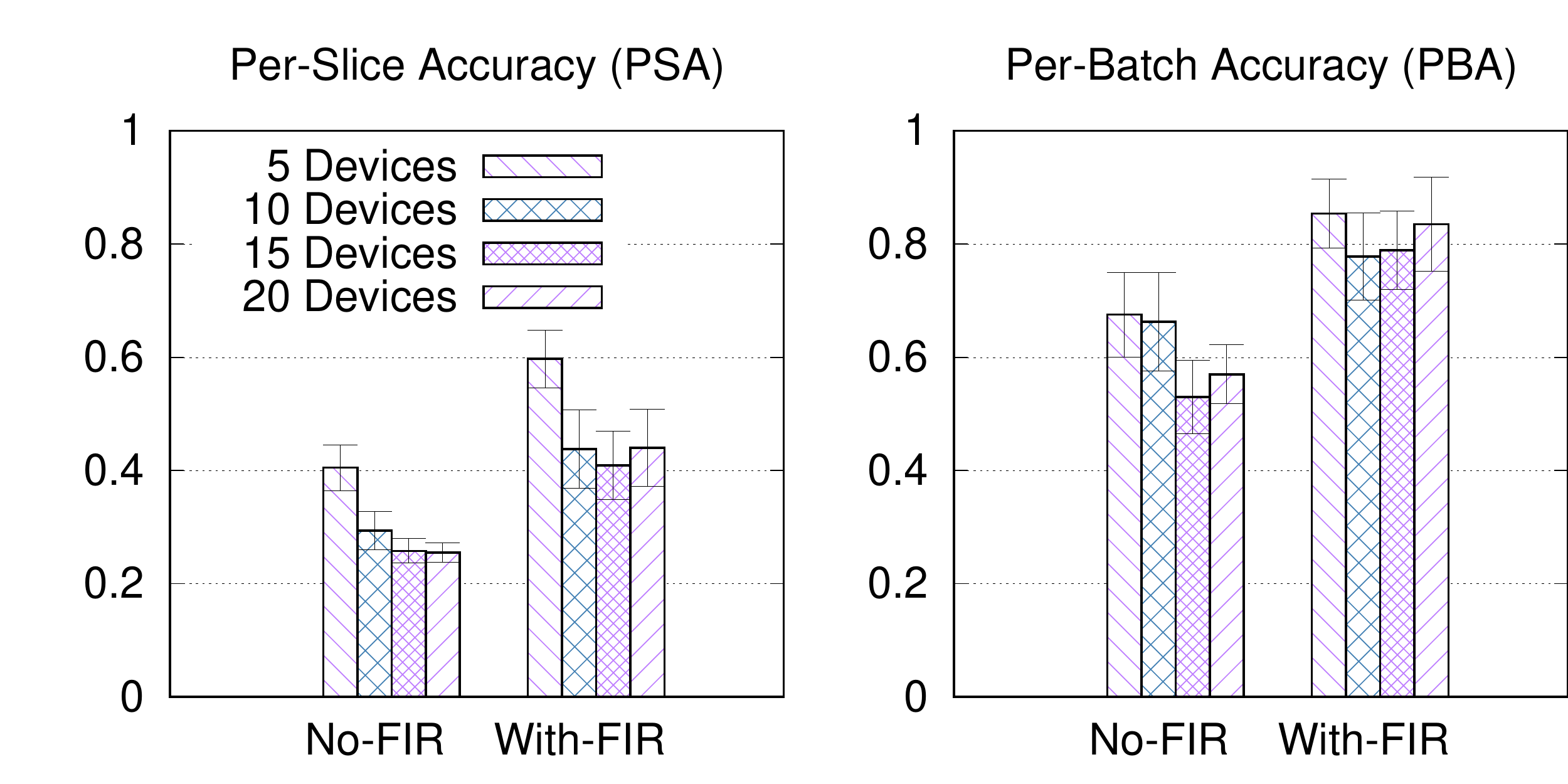}
    \caption{PSA and PBA, Experimental Testbed.\vspace{-0.6cm}}
    \label{fig:pba_testbed}
\end{figure}

Figure \ref{fig:pba_testbed} shows  the average PSA and PBA obtained by optimizing three devices over ten different recordings PSA as a function of the model size (\textit{i.e.}, 5 to 20 devices). We also show 95\% confidence intervals. The results obtained in Figure \ref{fig:pba_testbed} show that \textit{DeepFIR} is significantly effective in improving the PSA and PBA of our CNN-based fingerprinting system, which are improved by an average of about 57\% and 35\%, respectively. By accounting that the metrics are computed on a testbed of nominally-identically radios, we consider these results remarkable. 

\begin{figure}[!h]
    \centering
    \includegraphics[width=0.33\columnwidth,angle=-90]{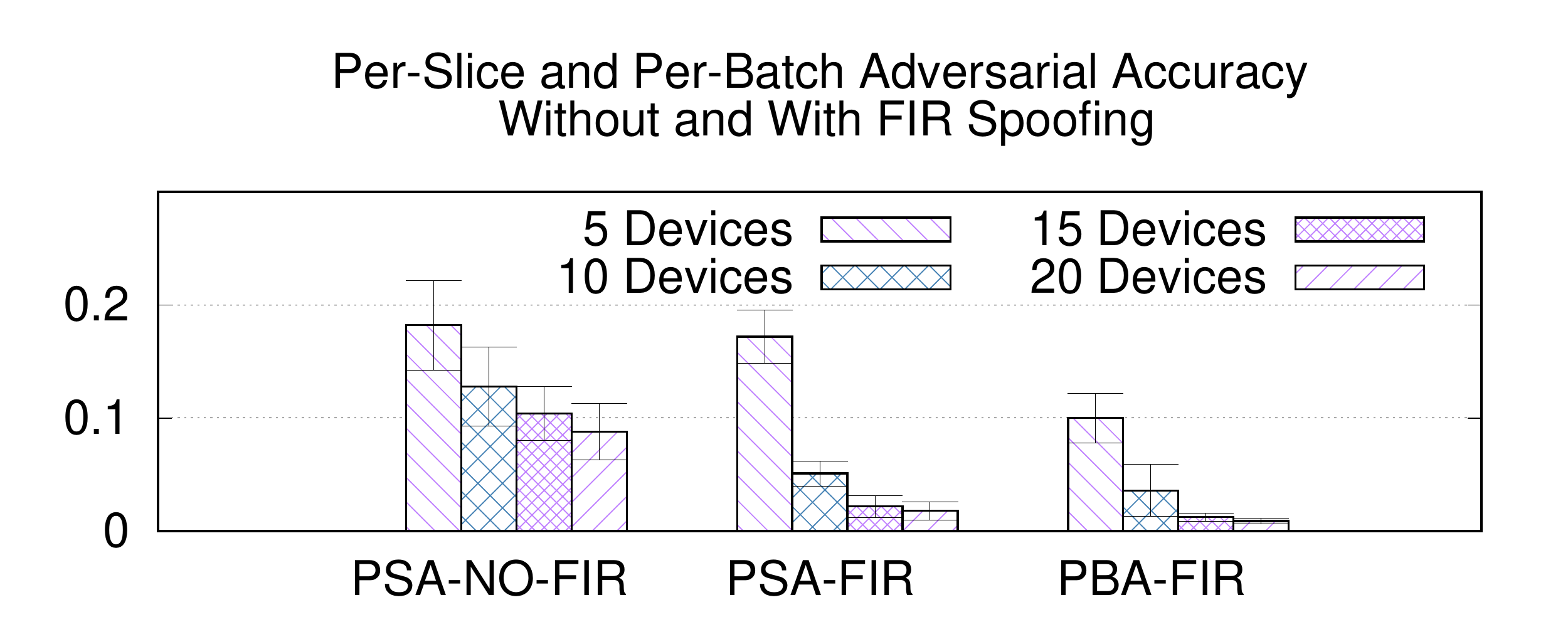}
    \caption{Adversarial Action, Experimental Testbed.\vspace{-0.5cm}}
    \label{fig:adv_testbed}
\end{figure}

The intuition of why PBA is lower than the PSA is that the FIR filter is optimized for a single slice only, thus its efficacy is expected to decrease as time passes and thus the wireless channel changes accordingly. Figure \ref{fig:adv_testbed} shows the PSA and the PBA obtained by an adversary trying to imitate another device's fingerprint by applying the same FIR. In these experiments, we fixed one device as adversary and used the FIR from other three devices to compute its PSA and PBA for each of its ten recordings. Figure \ref{fig:adv_testbed} shows that by using the legitimate device's FIR, the adversary transitions from an average PSA of 12\% to an average PSA of about 6\% (50\% decrease), corresponding to an average PBA of about 4\%. This ultimately confirms that since a FIR is optimized for a device's specific channel and impairments, an adversary cannot use a legitimate device's FIR to imitate its fingerprint.

\section{Fingerprinting Dataset Results}\label{sec:dataset_res}

We consider (i) a 500-device dataset of IEEE 802.11a/g (WiFi) transmissions;  and (ii) a 500-airplane dataset of Automatic Dependent Surveillance -- Broadcast (ADS-B) beacons, both obtained through the DARPA RFMLS program, and (iii) the RadioML 2018.01A dataset. \textbf{The dataset is publicly available for download at \url{http://deepsig.io/datasets}}. 

ADS-B is a surveillance transmission where an aircraft determines its position via satellite navigation and periodically broadcasts it at center frequency 1.090~GHz with sampling rate 1~MSPS and pulse position modulation. This makes ADS-B ideal to evaluate \emph{DeepFIR}'s performance on different channel/modulation scenarios. For both the WiFi and ADS-B datasets, data collection was performed ``in the wild'' (\textit{i.e.}, no controlled environment) with a Tektronix RSA operating at 200 MSPS. For the WiFi dataset, as in Section \ref{sec:opt_res} we demodulated the transmissions and trained our models on the derived I/Q samples. To demonstrate the generality of \emph{DeepFIR}, the ADSB model was instead trained on the unprocessed I/Q samples. 

\begin{figure}[!h]
    \centering
    \includegraphics[width=0.8\columnwidth]{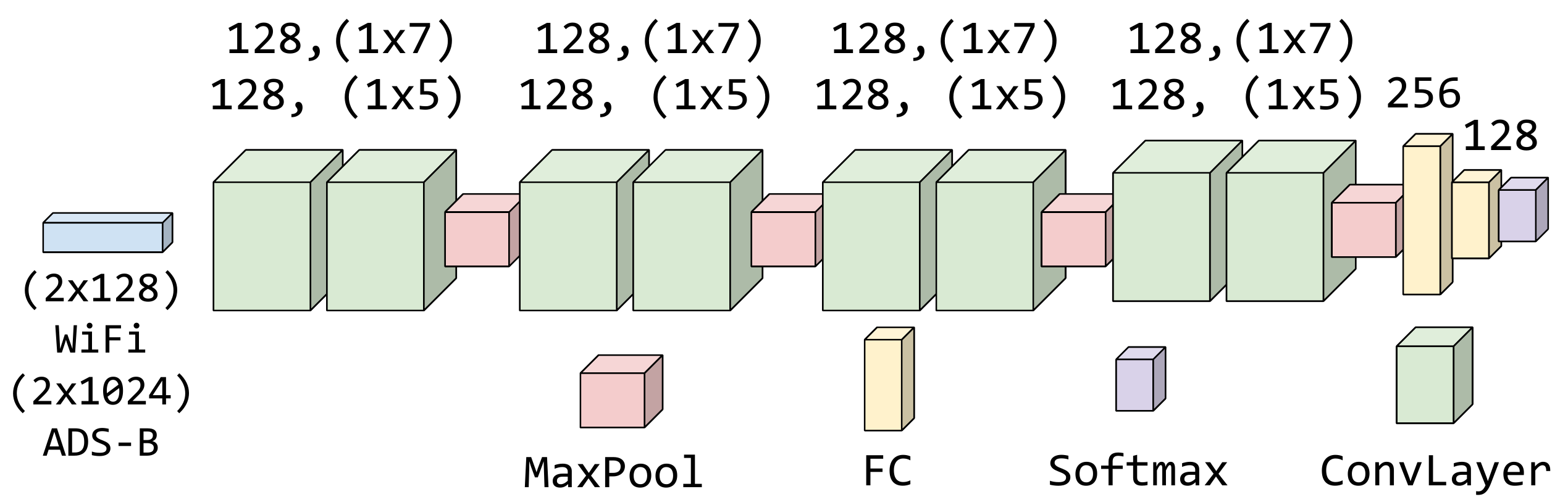}
    \caption{CNN used in our fingerprinting dataset experiments.\vspace{-0.3cm}}
    \label{fig:cnn_datasets}
\end{figure}

To handle the increased number of devices, we use the CNN architecture shown in Figure \ref{fig:cnn_datasets}. In the case of ADS-B we train our CNN on examples containing 1024 consecutive unprocessed I/Q samples. For WiFi, we only use 128 samples. Unless stated otherwise, PSA and PBA are computed on batches of 12 slices, each containing 25 inputs.

\begin{figure}[!h]
    \centering
    \includegraphics[width=0.8\columnwidth]{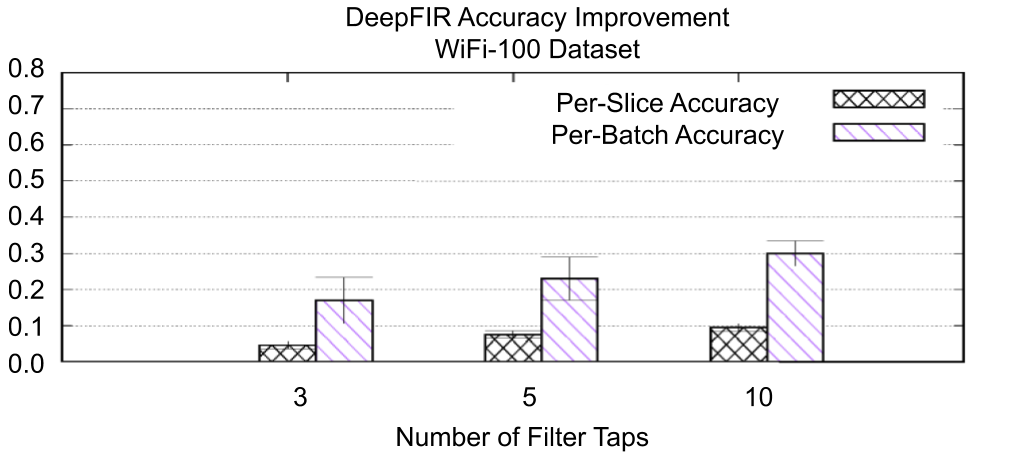}
    \caption{PSA/PBA Improvement, WiFi-100.\vspace{-0.6cm}}
    \label{fig:wifi100improvement}
\end{figure}

To compare our approach with the state of the art by Vo \emph{et al.}~\cite{12_vo2016fingerprinting}, which reports results on 93 devices, we trained our model on a subset of 100 devices, achieving PSA and PBA of .32 and .44, respectively. Figure \ref{fig:wifi100improvement} shows the PSA and PBA improvement brought by \textit{DeepFIR} FIR filtering as function of the number of FIR taps. In particular, the PBA improvement is around 30\% when 10 FIR taps are used, which brings the accuracy to about 74\% on the average, outperforming \cite{12_vo2016fingerprinting} which is 47\%.

\begin{figure}[!h]
    \centering
    \includegraphics[width=0.75\columnwidth]{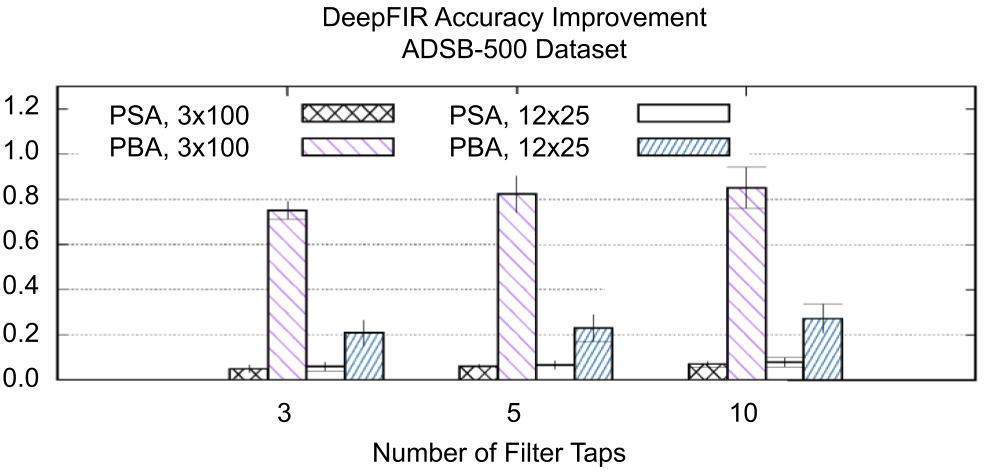}
    \caption{Per-Slice and Per-Batch Accuracy Improvement, ADSB-500, as a function of the number of inputs per slice (100 and 25) and the number of batches (3 and 12).\vspace{-0.3cm}}
    \label{fig:adsb500improvement}
\end{figure}

Figure \ref{fig:adsb500improvement} reports the PSA and PBA improvement on the ADS-B-500 dataset (baseline PSA and PBA of .5028 and .6193), which reach respectively about 10\%  and 50\% in the case of 10 filter taps. Figure \ref{fig:adsb500improvement} also shows that,  although the ADSB-500 PSA improvement is about the same experienced in the WiFi-100 model, the PBA improvement is significantly higher in ADSB-500 (30\% vs 50\%). This is thanks to the fact that a very small increase in PSA usually corresponds to a significant increase in PBA when the number of devices is higher. Also, ADS-B transmits in a less-crowded channel than WiFi, thus we expect our model and \textit{DeepFIR} to perform better on ADSB-500 when the number of inputs per slice increases.

\begin{figure}[!h]
    \centering
    \includegraphics[width=0.75\columnwidth]{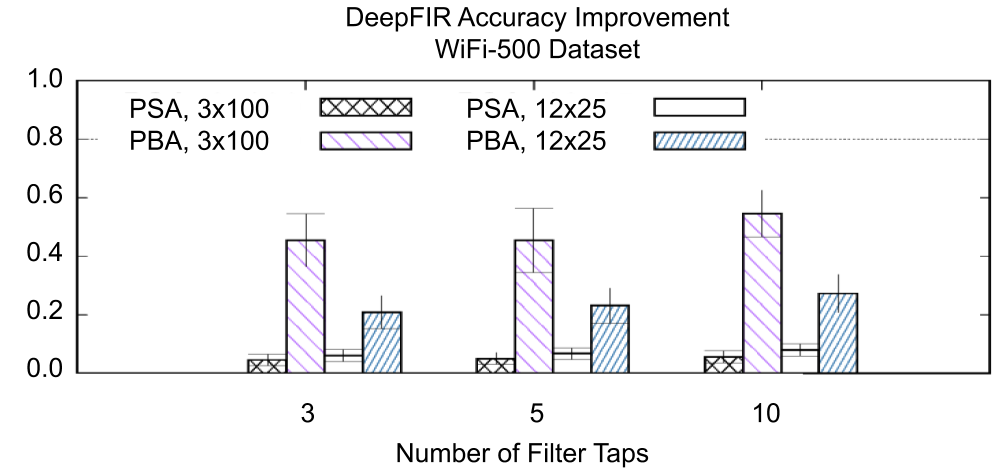}
    \caption{Per-Slice and Per-Batch Accuracy Improvement, WiFi-500, as a function of the number of inputs per slice (100 and 25) and the number of batches (3 and 12).\vspace{-0.3cm}}
    \label{fig:wifi500improvement}
\end{figure}

This is also confirmed by Figure \ref{fig:wifi500improvement}, where we show the PSA and PBA improvement in the WiFi-500 dataset as a function of the number of FIR taps and the number of inputs in each slice (25 and 100). Indeed, Figure \ref{fig:wifi500improvement} shows a similar PBA improvement when 10 FIR taps are considered. Furthermore, Figure \ref{fig:wifi500improvement} shows that the number of inputs per slice significantly impacts on the fingerprinting accuracy, especially on the PBA, in both ADSB-500 and WiFi-500. This is because (i) the FIR optimization increases in effectiveness as the number of inputs per slice increases, because the FIR is averaged over more channel realizations; and (ii) the boosting effect given by the PBA increases as the number of slices increases.

\begin{figure}[!h]
    \centering
    \includegraphics[width=0.75\columnwidth]{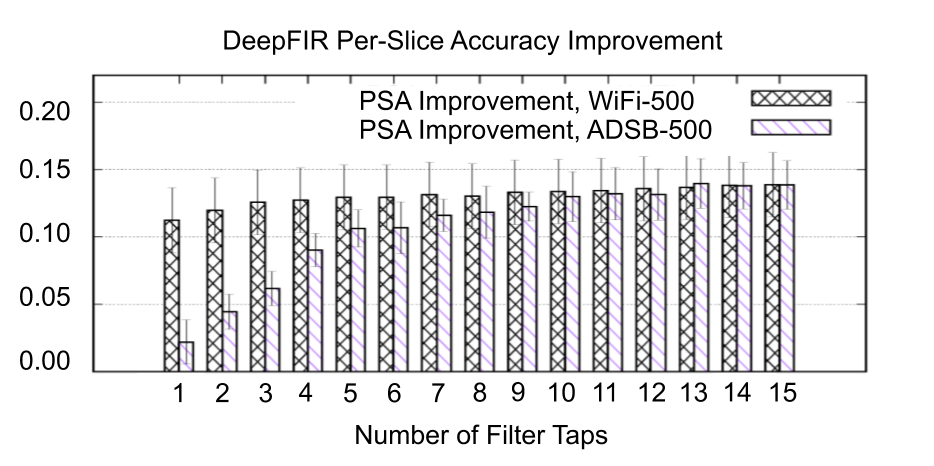}
    \caption{PSA Improvement as a function of the number of FIR taps. ADSB-500 converges slower than WiFi-500 given the FIR has to operate on unprocessed I/Q samples. \vspace{-0.3cm}}
    \label{fig:filtertapsadsb500}
\end{figure}

\begin{figure}[!h]
    \centering
    \includegraphics[width=0.4\columnwidth,angle=-90]{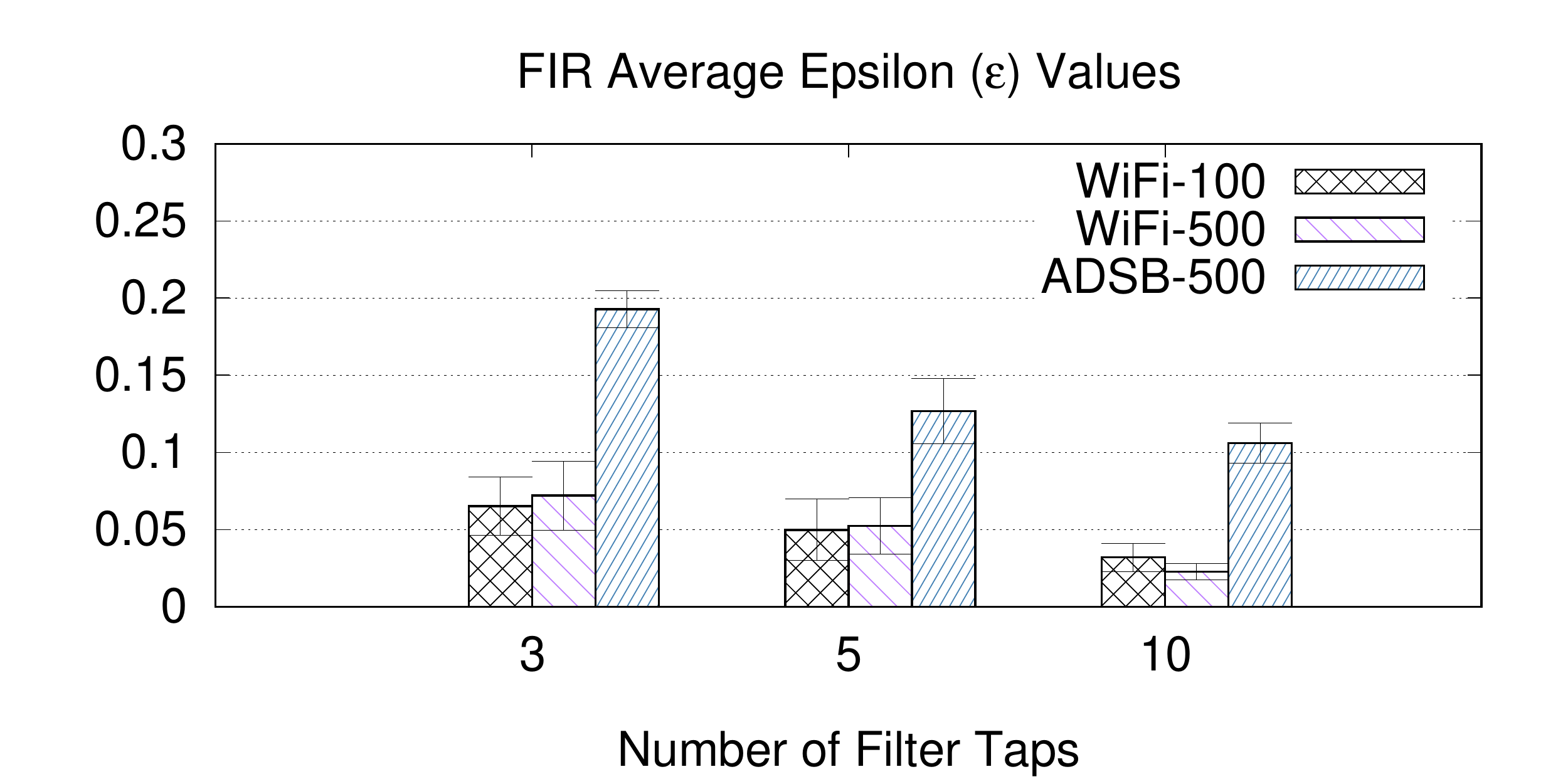}
    \caption{Average FIR Epsilon ($\epsilon$) values as a function of the number of FIR taps.\vspace{-0.8cm}}
    \label{fig:epsilon_taps}
\end{figure}

Figure \ref{fig:filtertapsadsb500} shows the PSA improvement for ADSB-500 and WiFi-500 as a function of the number of FIR taps. As we can see, the PSA improvement converges to approximately .15 as more FIR taps are used, in both cases. Very interestingly,  Figure \ref{fig:filtertapsadsb500} also shows that WiFi-500 converges more rapidly than ADSB-500. This is due to the fact that our ADSB-500 model was trained on unprocessed I/Q samples (\textit{i.e.}, without any channel equalization). Therefore, the number of FIR taps than \emph{DeepFIR} needs to obtain the same performance as in WiFi-500 increases, as the FIR has to compensate for a significant channel action in WiFi-500. Finally, Figure \ref{fig:epsilon_taps} depicts the average FIR $\epsilon$-value (defined in Section \ref{sec:ber}) as a function of the number of taps and the model considered. As anticipated in Section \ref{sec:ber}, Figure \ref{fig:epsilon_taps} concludes that the maximum $\epsilon$-value is below .2, which allows us to achieve low BER even when the FIR is applied. Interestingly enough, we notice that ADSB-500 requires taps that are on average higher than WiFi-100 and WiFi-500. This is due to the longer input (1024 vs 128 I/Q samples).\vspace{-0.5cm}

\section{Modulation Dataset Results}\label{sec:modulation_res}

In the following, we show the experimental results obtained on the RadioML 2018.01A dataset, which presents 24 different analog and digital modulations with different levels of signal-to-noise ratio (SNR) \cite{OShea-ieeejstsp2018}. For comparison purposes,  we consider the neural network in \cite{OShea-ieeejstsp2018}, Table III. This network is composed by 7 convolutional layers each followed by a MaxPool-2 layer, finally followed by 2 dense layers and 1 softmax layer. The dataset contains 2M examples, each 1024 I/Q samples long.

\begin{figure}[!h]
    \centering
    \includegraphics[width=\columnwidth]{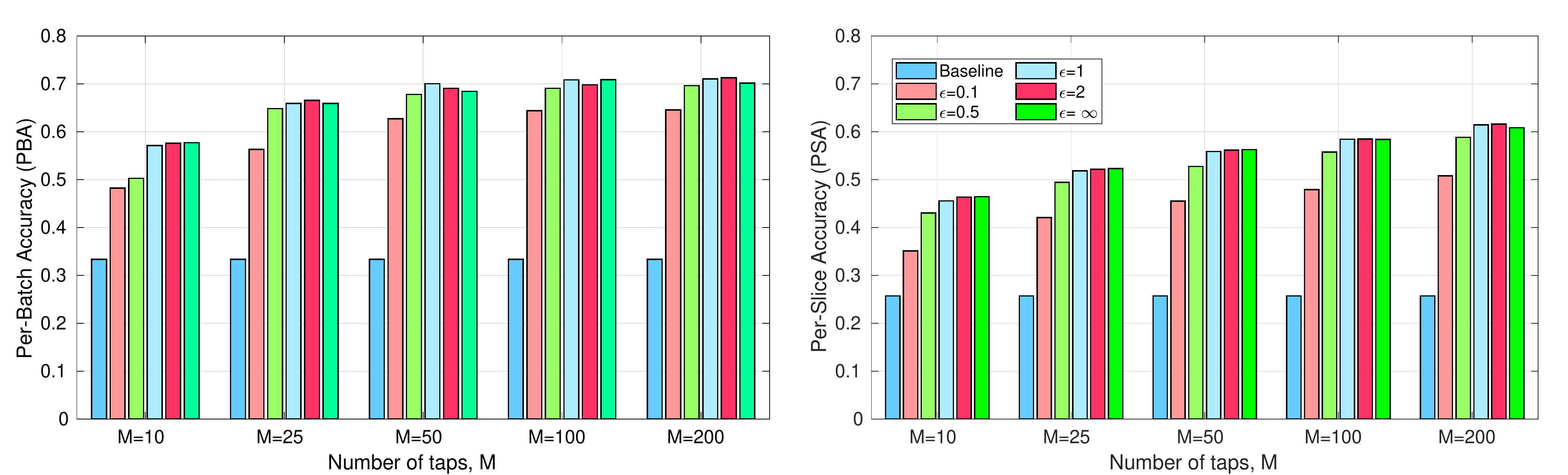}
    \caption{PSA and PBA as a function of $\epsilon$ for different number of FIR taps.\vspace{-0.3cm}}
    \label{fig:mod:accuracy}
\end{figure}

Figure~\ref{fig:mod:accuracy} shows the PBA and PSA with and without DeepFIR for a variety of different configurations of $\epsilon$ and number of FIR taps $M$. As expected, DeepFIR always improves both PBA and PSA in any configuration of $\epsilon$ and $M$. Similarly to prior results, the accuracy increases as the number of FIR taps increases until it reaches a plateau. We notice that both PSA and PBA increase when large values of $\epsilon$ are considered (\textit{e.g.}, the BER constraint is relaxed). In this case, DeepFIR focuses on maximizing the accuracy only disregarding the impact of FIR taps on the BER of the communications. However, even when $\epsilon=0.1$, the PBA obtained via DeepFIR is doubled when compared to the baseline model.

\begin{figure}[!h]
    \centering
    \includegraphics[width=\columnwidth]{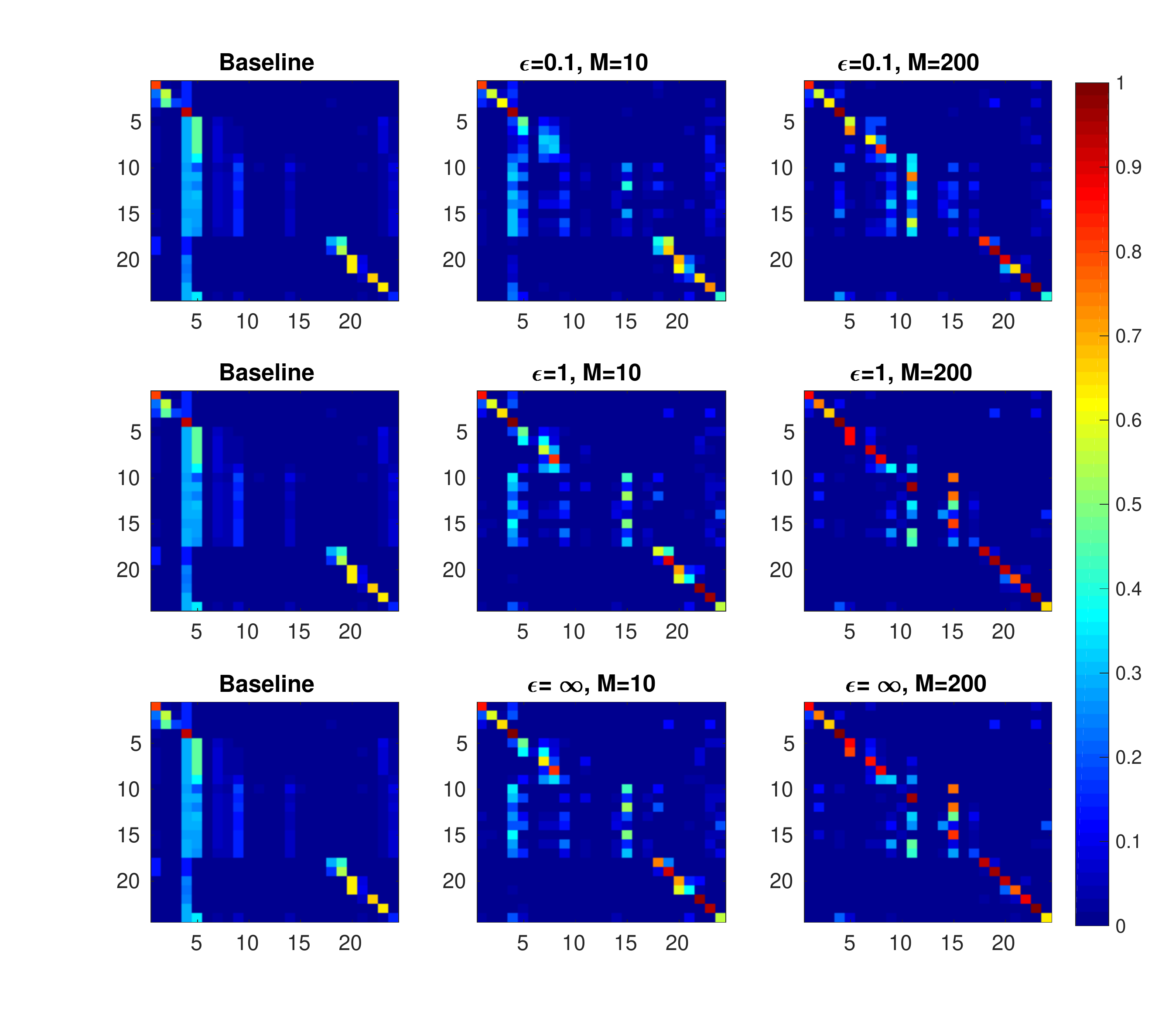}
    \caption{Confusion matrices for different $\epsilon$ values and number of FIR taps.\vspace{-0.3cm}}
    \label{fig:mod:confusion}
\end{figure}

To understand the impact of DeepFIR on the accuracy of the model, in Figure \ref{fig:mod:confusion} we show the confusion matrices for different DeepFIR configurations. While the baseline model misclassifies the majority of modulations, DeepFIR considerably improves the accuracy of the network. This is demonstrated by the presence of high values in the main diagonal of the confusion matrix. Although the best results are obtained when both $\epsilon$ and $M$ are large, the configuration with $\epsilon=0.1$ (\textit{i.e.}, DeepFIR computes FIR taps that can be successfully compensated at the receiver side, thus effectively limiting the impact of FIR filtering on the BER of the system), already achieves high accuracy (approximately 65\% PBA from Figure \ref{fig:mod:accuracy} when compared to 33\% PBA of the baseline model).

\begin{figure}[!h]
    \centering
    \includegraphics[width=\columnwidth]{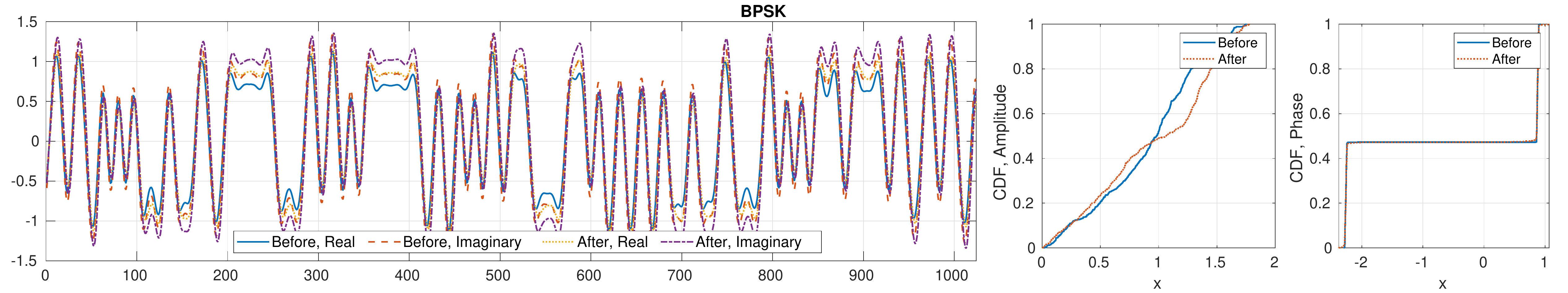}
    \caption{Impact of FIR filtering on waveforms and statistical distribution of I/Q samples for BPSK (Class 4) modulation.\vspace{-0.3cm}}
    \label{fig:mod:bpsk}
\end{figure}

\begin{figure}[!h]
    \centering
    \includegraphics[width=\columnwidth]{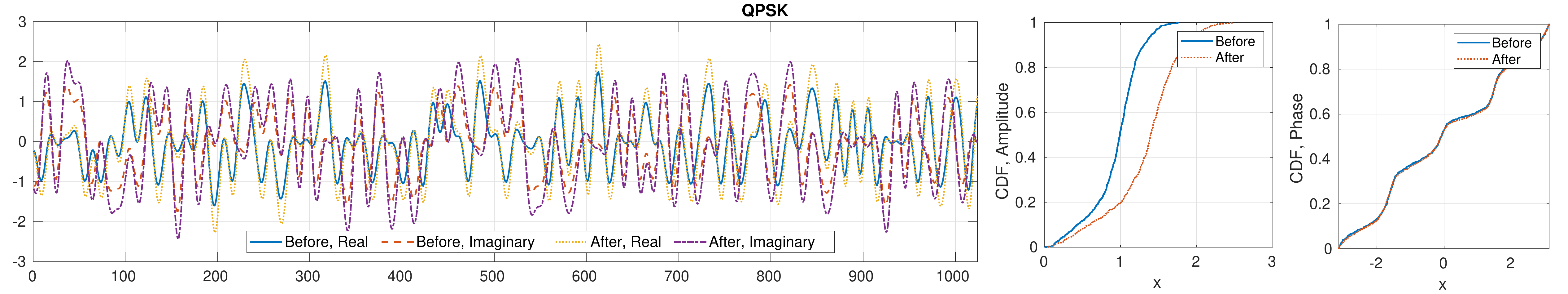}
    \caption{Impact of FIR filtering on waveforms and statistical distribution of I/Q samples for QPSK (Class 5) modulation.\vspace{-0.3cm}}
    \label{fig:mod:qpsk}
\end{figure}

\begin{figure}[!h]
    \centering
    \includegraphics[width=\columnwidth]{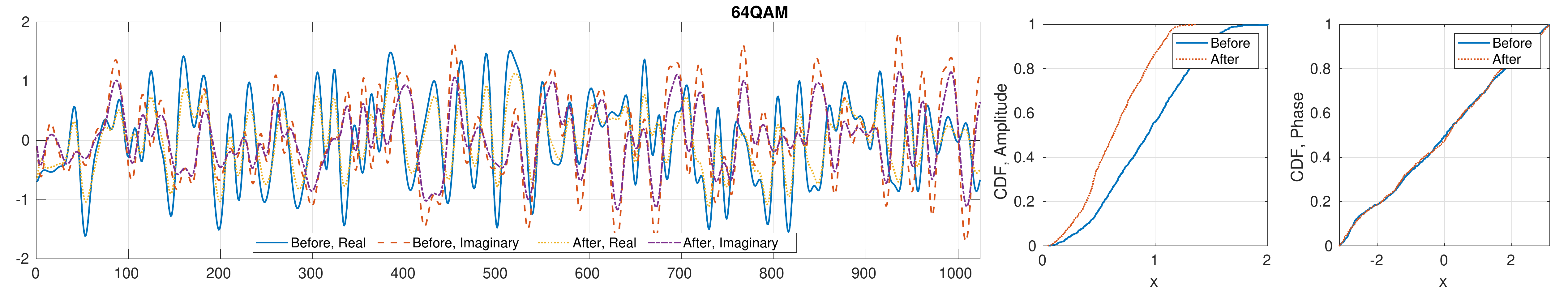}
    \caption{Impact of FIR filtering on waveforms and statistical distribution of I/Q samples for 64QAM (Class 15) modulation.\vspace{-0.3cm}}
    \label{fig:mod:qam}
\end{figure}

To better understand how DeepFIR impacts waveforms and displaces I/Q samples in the complex space, in Figs.~\ref{fig:mod:bpsk}, \ref{fig:mod:qpsk} and \ref{fig:mod:qam} we show waveforms before and after filtering for BPSK, QPSK and 64QAM modulations (i.e., classes 4,5 and 15), respectively. Unless otherwise stated, we consider $\epsilon=0.1$ and $M=10$. 

Figure \ref{fig:mod:bpsk} shows that FIR taps computed by DeepFIR are leaving the phase of the signal unchanged and are only minimally varying their amplitude. This is an interesting result as BPSK signals are demodulated by determining the phase of received I/Q samples. This way, a receiver would still be able to demodulate received BPSK waveforms properly without errors. The same also holds for QPSK modulated signals. As an example Figure~\ref{fig:mod:qpsk} shows that DeepFIR only affects the amplitude of transmitted I/Q samples while leaving the phase untouched (note the characteristic step CDF of the phase). 

Figure \ref{fig:mod:qam}, instead, shows the waveform for a 64QAM signal before and after DeepFIR filtering with $\epsilon=1$ and $M=10$. In this particular configuration, the two CDFs show that the impact of DeepFIR on the phase is slightly more relevant than in the case of PSK modulations, while the amplitude of the signal is reduce for most I/Q samples. Despite this reduction might seem destructive, recall that we can always compensate waveform alterations, especially for the considered setup with $\epsilon=1$ and $M=10$, meaning that the impact of DeepFIR on the demodulation procedure is negligible.  

\begin{figure}[!h]
    \centering
    \includegraphics[width=\columnwidth]{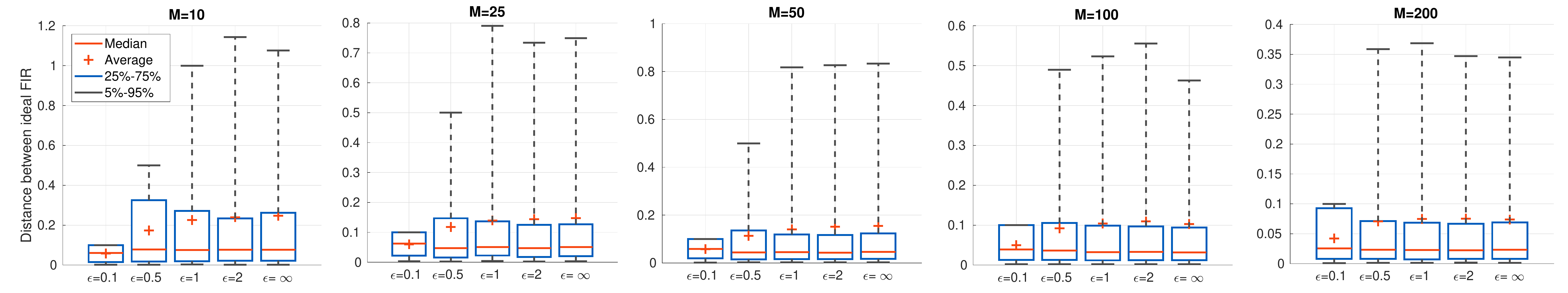}
    \caption{Statistical analysis of data-driven FIR taps.\vspace{-0.3cm}}
    \label{fig:mod:distance}
\end{figure}

Finally, in Figure \ref{fig:mod:distance} we present results on the statistical distribution of computed FIR taps. Recall that $\epsilon$ represents the maximum distance between FIR taps and the ideal FIR (i.e., $\boldsymbol{\phi}=(1,0,\dots,0)$). Figure~\ref{fig:mod:distance} shows that the data-driven approach (i) always satisfies the maximum distance constraint; and (ii) on average, the majority of taps shows a maximum distance from the ideal FIR less than 0.2. This is an important result because, as we have discussed in Section \ref{sec:ber}, the receiver can effectively compensate FIR action and recover the original transmitted waveform.

\section{Conclusions}

In this paper, we have proposed \emph{DeepFIR}, a methodology to make deep-learning-based  waveform classification algorithms channel-resilient. \emph{DeepFIR} leverages a finite input response filter (FIR) at the transmitter's side to compensate for current channel conditions, which is computed by the receiver through optimization and sent back as feedback to the transmitter. We have extensively evaluated \emph{DeepFIR} on a experimental testbed of 20 nominally-identical software-defined radios, as well as on datasets made up by WiFi and ADS-B transmissions and a 24-class modulation dataset. Experimental results have shown that \emph{DeepFIR} is significantly effective in increasing the classification accuracy of existing state-of-the-art deep learning models.

\section*{Acknowledgments}

This work is supported by the Defense Advanced Research Projects Agengy (DARPA) under RFMLS program contract N00164-18-R-WQ80. We are sincerely grateful to Paul Tilghman and Esko Jaska for their insightful comments and suggestions. The views and conclusions contained herein are those of the authors and should not be interpreted as necessarily representing the official policies or endorsements, either expressed or implied, of DARPA or the U.S. Government.

\footnotesize
\bibliographystyle{ieeetr}
\bibliography{main}

\end{document}